\documentclass[5p,times]{elsarticle}

\usepackage{graphics}
\usepackage{multirow}
\usepackage{float}
\usepackage{epstopdf}
\usepackage{booktabs}
\usepackage{color}
\usepackage{subfigure}
\usepackage{graphicx}
\usepackage{eqnarray}
\usepackage{lipsum}
\usepackage{lineno}
\usepackage{chngpage}
\usepackage{amsmath,amsfonts,amsthm,bm} 

\usepackage[amssymb,thinqspace,thinspace]{SIunits}

\modulolinenumbers[5]

\journal{journal}

\begin{document}

\begin{frontmatter}

\title{Ultra high-temperature deformation in a single crystal superalloy: Meso-scale process simulation and micro-mechanisms}

\author[addressmain]{Yuanbo T. Tang}
\author[addressfour]{Neil D'Souza}
\author[addressthree]{Bryan Roebuck}
\author[addressmain]{Phani Karamched}
\author[addressmain,addressfive]{Chinnapat Panwisawas}
\author[addresstwo]{David M. Collins}


\address[addressmain]{Department of Materials, University of Oxford, Parks Road, Oxford, OX1 3PH, United Kingdom}
\address[addressfour]{Rolls-Royce plc, PO Box 31, Derby, DE24 8BJ, United Kingdom}
\address[addressthree]{National Physical Laboratory, Hampton Road, Teddington, Middlesex, TW11 0LW, United Kingdom}

\address[addressfive]{School of Engineering, University of Leicester, University Road, Leicester, LE1 7RH , United Kingdom}
\address[addresstwo]{School of Metallurgy and Materials, University of Birmingham, Birmingham, B15 2TT, United Kingdom}


\begin{abstract}

A mesoscale study of a single crystal nickel-base superalloy subjected to an industrially relevant process simulation has revealed the complex interplay between microstructural development and the micromechanical behaviour. As sample gauge volumes were smaller than the length scale of the highly cored structure of the parent material from which they were produced, their subtle composition differences gave rise to differing work hardening rates, influenced by varying secondary dendrite arm spacings, $\gamma'$ phase solvus temperatures and a topologically inverted $\gamma/\gamma'$ microstructure. The $\gamma'$ precipitates possessed a characteristic `X' morphology, resulting from the simultaneously active solute transport mechanisms of thermally favoured octodendritic growth and N-type rafting, indicating creep-type mechanisms were prevalent. High resolution-electron backscatter diffraction (HR-EBSD) characterisation reveals deformation patterning that follows the $\gamma/\gamma'$ microstructure, with high geometrically necessary dislocation density fields localised to the $\gamma/\gamma'$ interfaces; Orowan looping is evidently the mechanism that mediated plasticity. Examination of the residual elastic stresses indicated the `X' $\gamma'$ precipitate morphology had significantly enhanced the deformation heterogeneity, resulting in stress states within the $\gamma$ channels that favour slip, and that encourage further growth of $\gamma'$ precipitate protrusions. The combination of such localised plasticity and residual stresses are considered to be critical in the formation of the recrystallisation defect in subsequent post-casting homogenisation heat treatments.

\end{abstract}

\begin{keyword}

\texttt \space  High temperature deformation \sep Single crystal \sep HR-EBSD \sep Micromechanics \sep ETMT \sep Superalloys
\end{keyword}

\end{frontmatter}

\section{Introduction}

Single crystal nickel-base superalloy turbine blades in modern aero-engines operate under harsh conditions, experiencing temperatures close to their melting point, high stresses that induce plastic deformation and surface degradation via oxidation \& corrosion mechanisms \cite{ReedSuperalloys}. Survival of these components has led to design strategies that provide active cooling; this includes thermal barrier coatings applied to the aerofoil surface, surface film cooling holes or intricate internal cooling passages. The latter presents the most difficult processing challenge, since it results in very complex internal cooling geometries with discontinuities in the component cross-section \cite{Dobbs1996}. During directional solidification, stress concentration features accumulate significant plastic strain, known to prevail in the temperature range below the incipient melting point, but well above the solvus temperature of the strengthening precipitate phase, $\gamma'$ \cite{LI20151}. The resultant plastic strain presents a significant problem during subsequent homogenisation heat treatments of the as-cast material as recrystallisation can occur \cite{Burgel2000}. In single crystal alloys, the formation of recrystallised grains is unacceptable as they are highly detrimental to both creep resistance \cite{Meng2010} and thermal-mechanical fatigue \cite{Moverare2009}. The economic impact of the resultant scrap rate motivates a mechanism informed understanding of recrystallisation that will guide developments in casting technology.

To uncover why recrystallisation occurs, a broad range of studies have been conducted to identify the influencing factors. The onset of recrystallisation in particular has have received a fair amount of attention, where it is agreed that there is a plastic strain threshold and a minimum critical temperature when this plastic strain is induced \cite{Cox2003,Panwisawas2013}. It is further noted that the process will prevail as result of a migrating recrystallised grain boundary, where the presence of $\gamma'$ precipitates will slow this process \cite{Porter1981}, where the $\gamma'$ precipitates are dissolved at the migrating recrystallisation boundary front \cite{Dahlen1980}. Similarly, a $\gamma/\gamma'$ eutectic is considered to act as pinning sites for advancing grain boundaries, limiting growth of recrystallised grains \cite{Wang2012}. Hence, the microstructure must play a key role in determining the susceptibility of an alloy to recrystallise. This has a direct relationship to the bulk alloy composition, which has been demonstrated to have some dependency on recrystallisation \cite{Mathur2014}. Finally, the dissolution of phases during recrystallisation is enhanced by the presence of residual stresses \cite{Jo2003}.

Both the in-service and processing behaviour of single crystal nickel-base superalloys has been traditionally derived from isothermal stress-strain measurements, often performed under constant stress or constant strain rate \cite{DUPEUX19872203,WANG20131179,SVOBODA1997125,PREUNER2009973}. The transient temperature regimes subjected to nickel-base superalloys during casting may be better represented by non-isothermal creep deformation experiments, where there is extensive data in the high temperature (up to $\sim$1200\,$^\circ$C), low stress condition in several alloy systems including CMSX4 \cite{Giraud2013,Schwalbe2018}, MC2 \cite{CORMIER20076250,CORMIER20106300,Cormier2010} \& MC-NG \cite{LEGRAVEREND201455,LEGRAVEREND2014990}. In creep behaviour, capturing the thermo-mechanical history dependence of deformation and microstructure is known to be critical in predicting subsequent plasticity \cite{MA20081657,ZHU20124888}, which must also play a governing role in the behaviour of transient micromechanical and microstructural phenomena during an investment casting process. This is further supported by measurements obtained during cooling experiments from casting relevant temperatures where stress relaxation, $\gamma/\gamma'$ stress partitioning and work hardening including dislocation structures are shown to be highly dependent on the microstructral history during cooling \& creep experiments on CMSX-4 \cite{Panwisawas2017, Panwisawas2018}. Furthermore, the degree to which these properties factors vary and subsequently control microscopy behaviour is location dependent within the cast structure, due to micro-segregation remnant from the solidification \cite{DSOUZA2020138862}.

In spite of extensive work already performed to date, understanding the mechanisms that lead to recrystallisation in single crystal nickel-base superalloys has not yet reached a consensus. Reasons behind may be threefold. Firstly, there exists a lack of thermo-mechanical simulation data with suitable fidelity to conditions experienced by the alloy during casting, in particular to replicate the extreme temperatures ($\sim$1300~$^{\circ}$C) and time scales of the process. In addition, the site/history dependency on deformation is seldom reported -- conventional testing designs typically comprise a large sample volume, which averages out the contribution of individual dendrites. Furthermore, the micromechanisms that are operative under these conditions remained unclear due to inadequacy of direct evidence. Therefore, it has proven challenging to predict, control and prevent the formation of secondary grains in the susceptible casting geometries.

This study utilises high resolution electron backscatter diffraction (HR-EBSD) datasets to newly understand the development of the stress states and dislocation structures relevant to ultra-high temperature deformation, at length scales finer than the $\gamma/\gamma'$ microstructure. This was performed on a single crystal nickel-base superalloy subjected to the macroscopic stress states considered most critical for mechanisms relevant to recrystallisation. This has been possible via the design of an industrially-relevant process simulation using an electro-thermal mechanical testing (ETMT) system with miniaturised samples. The simulation replicates the thermo-mechanical environment subjected to alloys in the solid state, that proceeds soon after solidification in the investment casting process, as the alloy cools from the eutectic temperature to just below the $\gamma'$ solvus temperature. The resultant structure was preserved by cooling rapidly, followed by post-mortem characterisation, enabling the dynamic interplay between the governing micromechanics and microstructure to be revealed.

\section{Experimental Methodology} 

\subsection{Electro-thermal mechanical testing (ETMT)}

This study focuses on the thermo-mechanical behaviour of the single crystal Ni-base superalloy CMSX-4, as the temperatures falls in the solid-state from the eutectic temperature and through the $\gamma'$ solvus, with deformation simultaneously applied. This procedure represents an ultra-high temperature region of an industrial casting process; gaining knowledge here is deemed technologically critical for optimised control of microstructure whilst avoiding casting defects. Experiments were performed on single crystal cylinders of the CMSX-4 alloy, of nominal composition shown in Table \ref{Tab1}, which were directionally solidified with the $\langle 001 \rangle$ direction of the cast material to be within 10 $^{\circ}$ of the cylinder axis. Each cylinder measured $\sim$60 mm in length and $\sim$10\,mm in diameter. From these as-cast bars, matchstick shaped tensile specimens were electro-discharge machined (EDM) to have a 40\,mm   length and a cross sectional area slightly larger than 2\,mm\ $\times$\ 1\,mm. Any re-cast layer remnant on the surface following machining was removed with abrasive media. The process simulation was performed on an ETMT system, where user defined thermal and mechanical profiles can be applied to minaturised test specimens. Specimens were heated via a DC electric current that were held between water cooled grips, with macroscopic strain calculated from the measured changes in electrical resistance. This measurement has an uncertainty that gives a strain resolution of 0.05\,\%. Deformation was subjected to samples through an in-line drive, displacement controlled 0.5\,kN mechanical loading assembly. Full details of this method can be found elsewhere \cite{Roebuck2001,Roebuck2004,Sulzer2004}. 

\begin{figure*}
		\centering
		\includegraphics[width=0.8\textwidth]{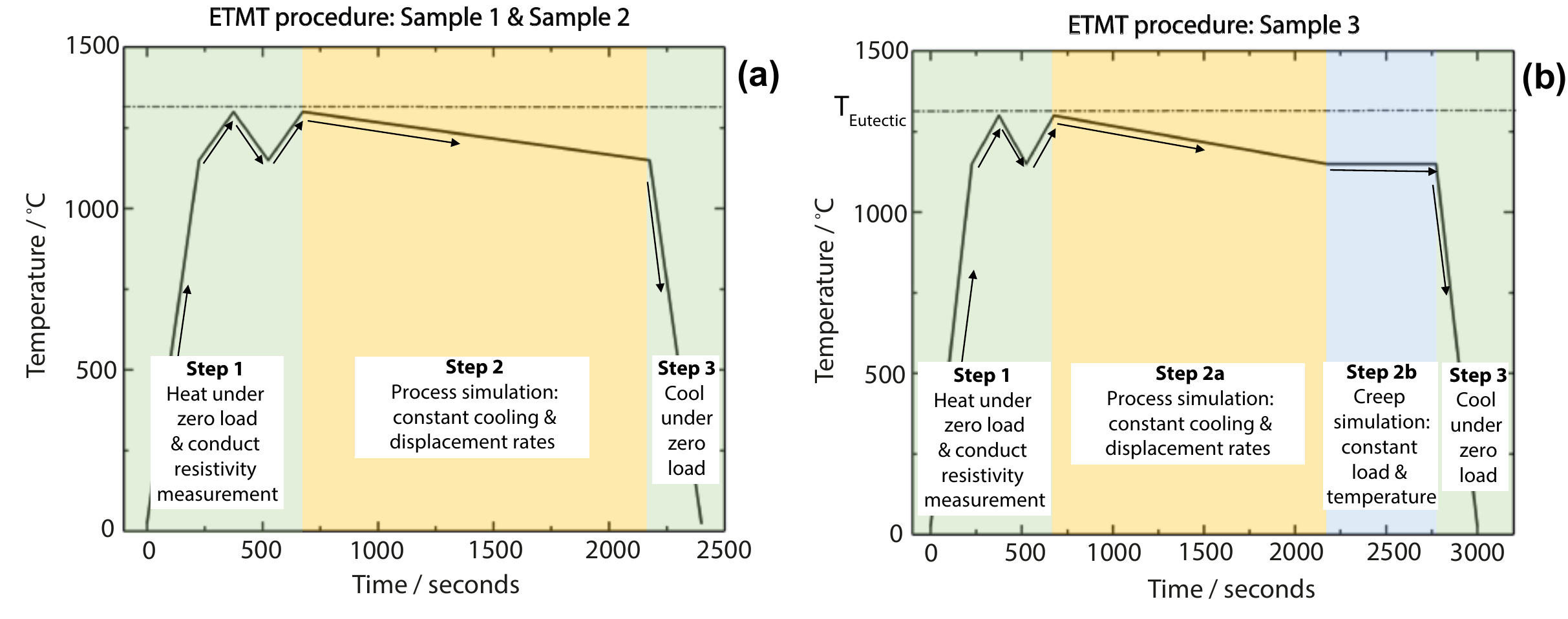}
		\caption{Temperature profiles applied to casting simulation samples using an electro-thermal mechanical testing machine. The temperature and mechanical deformation details at each step is summarised in Table 2.}
		\label{Method_ETMT}
	\end{figure*}

Samples of CMSX-4 were subjected to process simulations with deformation and thermal profiles illustrated schematically in Fig. \ref{Method_ETMT}. The strain measurements were initially calibrated in `Step 1' from resistivity measurements obtained from a short heating and cooling cylce under zero load in the vicinity of the eutectic temperature, $T_{\rm Eutectic}$. The process simulation (Step 2 \& Step 2a) comprised each sample being cooled at 0.1\,$^{\circ}$C/s, which is typical for an industrial casting processing, under strain rates, $\dot{\varepsilon}$ , stated in Table \ref{Tab2}. The strain rates examined were selected to mimic the deformation of a real process that arises from the thermal mismatch between the ceramic mould and the metal. Sample 3 was initially deformed more rapidly than Samples 1 \& 2, but additionally includes a short load dwell (Step 2b) to exaggerate an extreme scenario where the cooling rate is low but deformation still prevails. Once samples had reached 1150\,$^{\circ}$C, they were cooled rapidly under zero load (Step 3) to preserve their elevated temperature microstructures and deformation structures. Each sample was subsequently characterised post-mortem.

\begin{table*}[h]
\centering
\caption{Nominal composition of CMSX-4 measured using ICP-OES, ICP-combustion (Carbon) and Spark OES (Boron).}
\label{Tab1}
\begin{tabular}{l*{12}{c}r}
\hline \hline
 Element &  Ni   &  Cr    &  Co    &  Al  &  Ti    &   Ta    &  W     &  Mo    &  Hf   &  Re  \\
\hline
wt.\% & Bal   & 6.5 &  9.6    & 5.6  &  1.0   &  6.5   &  6.4   &  0.6  &  0.1  & 3.0  \\
\hline

\hline \hline
\end{tabular}
\end{table*}

	
 \begin{table*}[h]
 \begin{adjustwidth}{-1in}{-1in}  
	\centering
	\caption{Temperature, displacement and load control parameters sued for sample 1-3 during casting simulation.}
	\label{Tab2}
	\begin{tabular}{c*{9}{c}r}
		\hline \hline
		 & Step 1 & {Step 2/2a} & Step 2b & Step 3  \\
		\hline
		& Resistivity measurement    & {Casting simulation} & Creep  & Cooling  \\
		\hline
	\multirow{2}{*}{Sample 1 \& Sample 2}& 25$\rightarrow$ 1300 $\rightarrow$1150$\rightarrow$ 1300\,[$^{\circ}$C]  & 1300$\rightarrow$1150\,[$^{\circ}$C] at 0.1\,$^{\circ}$C\,s$^{-1}$  &\multirow{2}{*}{N/A} & 1150$\rightarrow$25~[$^{\circ}$C]\\
	& zero load & $\dot{\varepsilon} = \sim2$$\times10^{-5}$  &  &  zero load \\
		
	\hline 
		\multirow{2}{*}{Sample 3}& 25$\rightarrow$1300$\rightarrow$1150$\rightarrow$1300 [$^{\circ}$C]  & 1300$\rightarrow$1200~[$^{\circ}$C] at 0.1\,$^{\circ}$C\,s$^{-1}$ & 1150\,$^{\circ}$C & 1150$\rightarrow$25~[$^{\circ}$C] \\
	& zero load  &  $\dot{\varepsilon} = \sim3.5\times10^{-5}$ & 88\,MPa &  zero load \\
		
		\hline \hline

	\end{tabular}
	\end{adjustwidth}
\end{table*}

\subsection{Electron Microscopy}

Following the casting simulation experiments, the ETMT specimens were mounted in a conductive resin and polished using abrasive media and finished with a 0.04\,\micro m colloidal silica for 3 minutes prior to electron microscopy. For micrographs obtained, the $\gamma/\gamma'$ microstructure including the eutectic phase could be observed following electrolytic etching with a 10\,\% H$_3$PO$_4$ aqueous solution at 3\,V. This etchant preferentially attacks the $\gamma$ phase, leaving the $\gamma'$ phase in relief.

A Zeiss merlin field emission gun scanning electron microscope (FEG-SEM) was utilised to study the samples. The microscope was equipped with an angular selective backscatter electron (AsB/BSE) detector and an immersion lens secondary electron (In-lens/SE) to obtain both compositional and topological contrast images. A Bruker e$^-$Flash$^{\rm HR}$ EBSD detector operated with Esprit 2.0 software was also used; EBSD diffraction patterns (EBSPs) were collected at 800\,$\times$\,600 resolution in 16 bits and were stored for offline analysis. Data were acquired with the microscope operating with a 5\,nA probe current and a 20\,kV electron beam energy. EBSD maps were acquired from each sample at two magnifications, with dimensions $23\times17$\,\micro m$^2$ \& $92\times68$\,\micro m$^2$, with step sizes of 45\,nm  \& 111\,nm, respectively. 

HR-EBSD was used to measure the lattice rotations and geometrically necessary dislocation density field, as well as residual elastic stresses in CMSX-4 samples at a sub-micron length scale. This method employs a cross-correlation based analysis of diffraction patterns to measure subtle changes in the geometry of the crystal, of misorientaion sensitivity of $6\times10^{-3}$\,$^{\circ}$ \cite{WILKINSON2006307} of up to $\sim$11\,$^\circ$ of crystal rotation \cite{BRITTON20111395} and strain sensitivity of at least $1\times10^{-4}$ \cite{Villert2009}. This method is reliant on a reference pattern selected within a grain and image shifts are measured via a cross correlation function between the reference pattern and test patterns within this grain. As samples here were single crystals, one reference pattern was obtained with the numerically highest pattern quality. It was assumed that comparing a selected reference to any point within a $\gamma$ or $\gamma'$ phase was valid as the CMSX-4 has near identical $\gamma$ \& $\gamma'$ lattice parameters at room temperature \cite{Panwisawas2017}. From the measured diffraction pattern shifts, a deformation gradient tensor is defined, from which strain and rotation components can be derived using a finite decomposition framework \cite{Britton_2018}. The reader is referred elsewhere for a fuller description of the HR-EBSD method  \cite{WILKINSON2012366,Britton2013} and details of its mathematical basis \cite{WMG2006_1,WILKINSON2006307}. The measured lattice curvature can be related to the GND density \cite{ARSENLIS19991597} by solving Nye's dislocation tensor \cite{Nye}; here GND density fields were estimated from the EBSD measurements \cite{PANTLEON2008994,Wilkinson2010}. The results obtained correspond to the total dislocation density from all 18 types of dislocation with a $\langle110\rangle$ Burgers vector for an FCC crystal structure, thus all dislocations are assumed to be perfect for both the $\gamma$ and $\gamma'$ phases.

\subsection{Differential Scanning Calorimetry (DSC)}

To identify the phase transition reactions of the material, differential scanning calorimetry measurements were performed on sections of the non-gauge regions of the ETMT samples. Each disc-shaped sample had a $\sim$40\,mg mass and a flat surface that was ground with abrasive SiC paper to 4000 grit, then ultrasonically cleaned in ethanol prior to testing. The specimen was heated in an platinum crucible with alumina liner to 1450\,$^{\circ}$C at 10\,$^{\circ}$C/min with a constant argon flow of 20 ml/min to prevent oxidation. After holding at this temperature for 15 mins, the samples were cooled at 10\,$^{\circ}$C/min. The data from each sample were corrected by the heat flow trace of an empty crucible that was subjected to the same thermal cycle.

\section{Results}

\subsection{Macroscopic behaviour}

	\begin{figure*}
		\centering
		\includegraphics[width=0.8\textwidth]{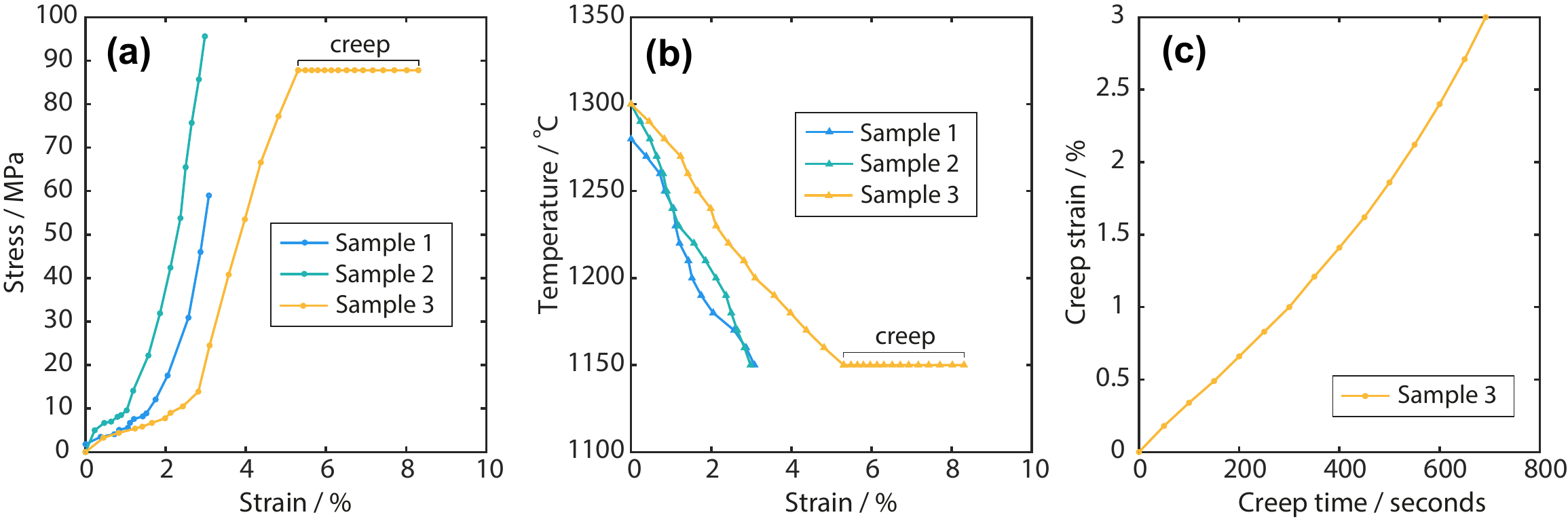}
		\caption{During the ETMT process simulations, the stress-strain response for Samples 1-3 were measured (a), as well as the corresponding temperature profile (b). For Sample 3, creep strain data was also obtained (c).}
		\label{Fig_stress_strain}
	\end{figure*}
	
CMSX-4 specimens were subjected to 3 thermal-loading cycles, denoted hereon as Sample 1, 2 \& 3. As shown in Fig. \ref{Fig_stress_strain}a, Samples 1 \& 2, were heated (Step 1), followed by the casting simulation (Step 2), and were rapidly cooled (Step 3) under no load to preserve the microstructure, whilst Sample 3 had the same casting simulation (Step 2a) plus a load dwell (Step 2b) to investigate the additional effect of creep (as experienced by real castings, under certain processing conditions), as shown in Fig. \ref{Fig_stress_strain}b. The corresponding stress-strain response during the three casting simulations are shown in Fig. \ref{Fig_stress_strain}c. Sample 3 was additional held at 1150$^\circ$C at 88\,MPa for $\sim$20\,mins where an additional 3\,$\%$ plastic strain was accumulated. It is clear that the strain hardening rate of each sample was drastically different. Samples 1 \& 2 reached equivalent plastic strains, but experienced very different strain hardening rates, leading to quite different resultant stresses. Sample 3 obtained the largest plastic deformation, expected from its higher strain rate than Sample 1 \& Sample 2, however, it unexpectedly has the lowest initial strain hardening rate (for strains $<$ 3\,\%) of all samples. A sharp increase in strain hardening rates is observed at 1.0\%, 1.8\% and 2.8\% strains, corresponding to $\sim$1240\,$^\circ$C, $\sim$1220\,$^\circ$C \& $\sim$1210\,$^\circ$C, in Samples 1, 2 \& 3, respectively. These points indicates the formation temperatures of the $\gamma'$ precipitates for each sample as they cool.
	
\begin{figure}
		\centering
		\includegraphics[width=\columnwidth]{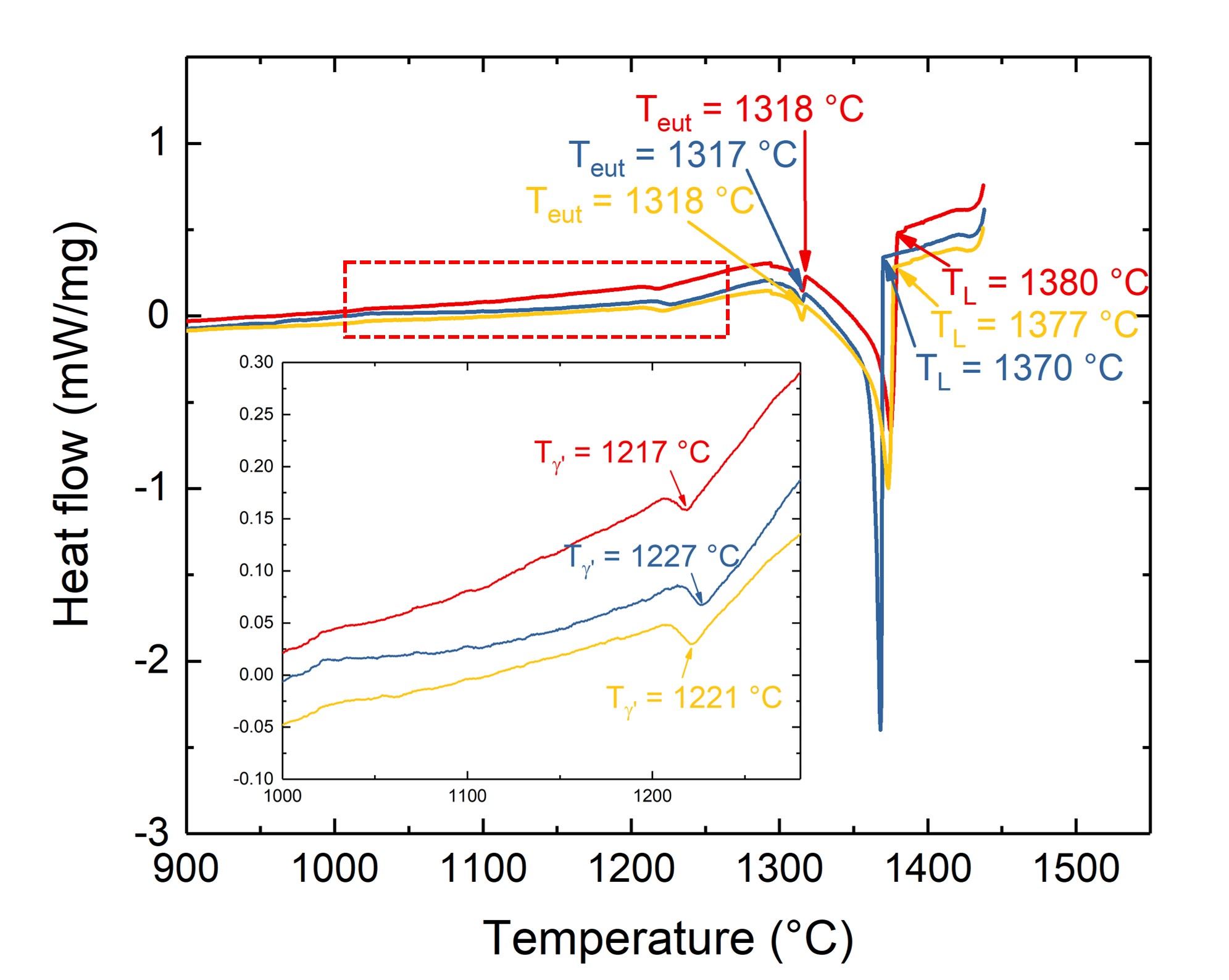}
		\caption{Differential scanning calorimetry cooling measurements of three specimens tested from an as-cast initial condition. The liquidus temperatures, $T_{\rm{L}}$, eutectic transformation temperatures, $T_{\rm{eut}}$, and the $\gamma'$ precipitate solvus temperatures, $T_{\gamma'}$, are labelled for each specimen.}
		\label{Figure_DSC}
	\end{figure}

To further corroborate the observed variation in macroscopic deformation behaviour within an as-cast bar, the repeatability of the phase transformation temperatures was explored via DSC measurements. These DSC measurements were obtained from volumes that were similar to the gauge volume of the ETMT test specimens, and were therefore deemed to be a suitable methods to correlate. Cooling curves for 3 samples, each taken from random as-cast bar locations are shown in Fig. \ref{Figure_DSC}. For these samples, the eutectic temperature showed little sample to sample variation at $\sim$1318$^{\circ}$C, whilst the liquidus and $\gamma'$ solvus temperatures were found to vary by $\pm5^{\circ}$C.

\subsection{Microstructure evolution during casting simulation}
 
\begin{figure*}
		\centering
		\includegraphics[width=0.9\textwidth]{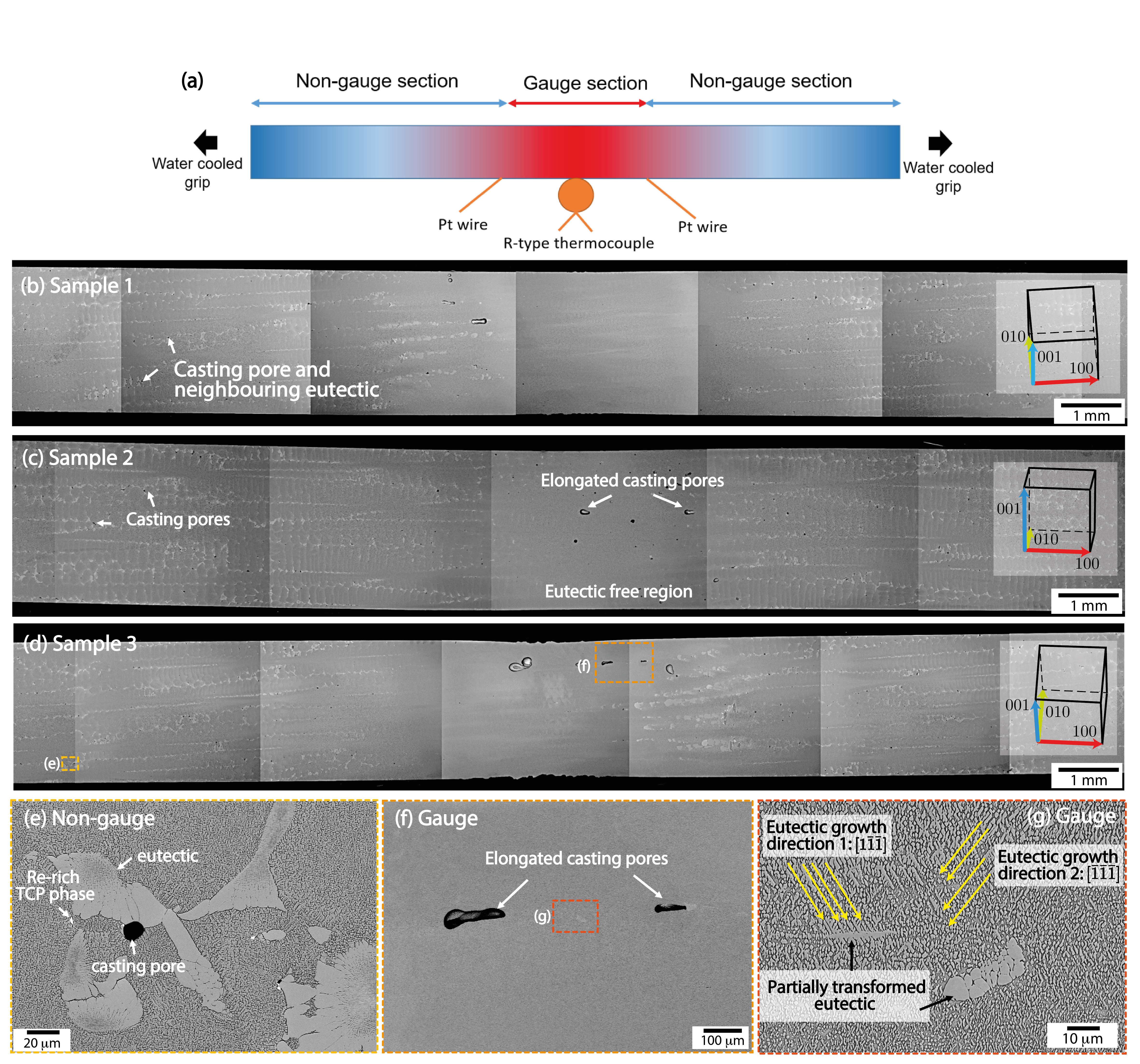}
		\caption{Features of the microstructures are identified along the length of the ETMT test specimens; each specimen is classified into regions  illustrated in (a) from the parabolic temperature profile of the test specimens including a `gauge section' where the sample temperature was high, and a water cooled `non-gauge region'. Stitched SE images along the length of ETMT specimens are shown for Samples 1-3 in (b-d), respectively. Magnified views of the non-gauge region show eutectic, casting pores and TCP phases (e), elongated pores within the gauge (f), and transformed eutectic within the gauge (g).}
		\label{Fig_microstructures}
	\end{figure*}

A schematic diagram of an ETMT specimen is shown in Fig. \ref{Fig_microstructures}a, where results were obtained from specimens heated via a direct current and were displaced by water cooled grips. Due to this cooling, the samples experience a parabolic temperature profile, with the a region spanning $\sim$4-5\,mm (measured at 800$^\circ$C-900$^\circ$C) at the specimen centre that is subjected to a steady-state temperature \cite{Roebuck2004}, controlled by an R-type thermocouple; this region is considered to be the specimen gauge section. It is noted that at higher temperatures, particularly above 1200$^\circ$C, the temperature profile becomes more uniform. Nonetheless, the variance in position specific thermal profile along the length of the specimens is evident from the BSE micrographs obtained from the specimen surface, as shown in Fig. \ref{Fig_microstructures}b - Fig. \ref{Fig_microstructures}d.

The microstructure of the three specimens, Fig. \ref{Fig_microstructures}b - Fig. \ref{Fig_microstructures}d, are broadly similar. The dendritic structures remnant from solidification in the non-gauge regions of the specimen are visible; primary and secondary arms can be seen. The significant segregation during constitutional undercooling results in regions of eutectic, often associated with enriched regions of aluminium, tantalum and titanium \cite{Fuchs2001}, here shown (Fig. \ref{Fig_microstructures}e) to be present in tens/hundreds of $\micro$m size islands that neighbour dendritic and interdendritic regions, consistent with typical as-cast CMSX-4 microstructures \cite{Wilson2003}. Here, the eutectic phases were seen with circular casting pores and neighbouring Re-rich TCP phases. Comparatively, the microstructure in the gauge region (Fig. \ref{Fig_microstructures}f and Fig. \ref{Fig_microstructures}g) is quite different, appearing more homogenous with fewer distinct dendritic structures. This region does, however, show numerous casting pores that are elongated in the tensile direction. Interestingly, the partially transformed eutectic phase in Fig \ref{Fig_microstructures}g illustrates preferential growth in the $\langle1 1 1\rangle$ directions. 

\begin{figure*}
		\centering
		\includegraphics[width=1.0\textwidth]{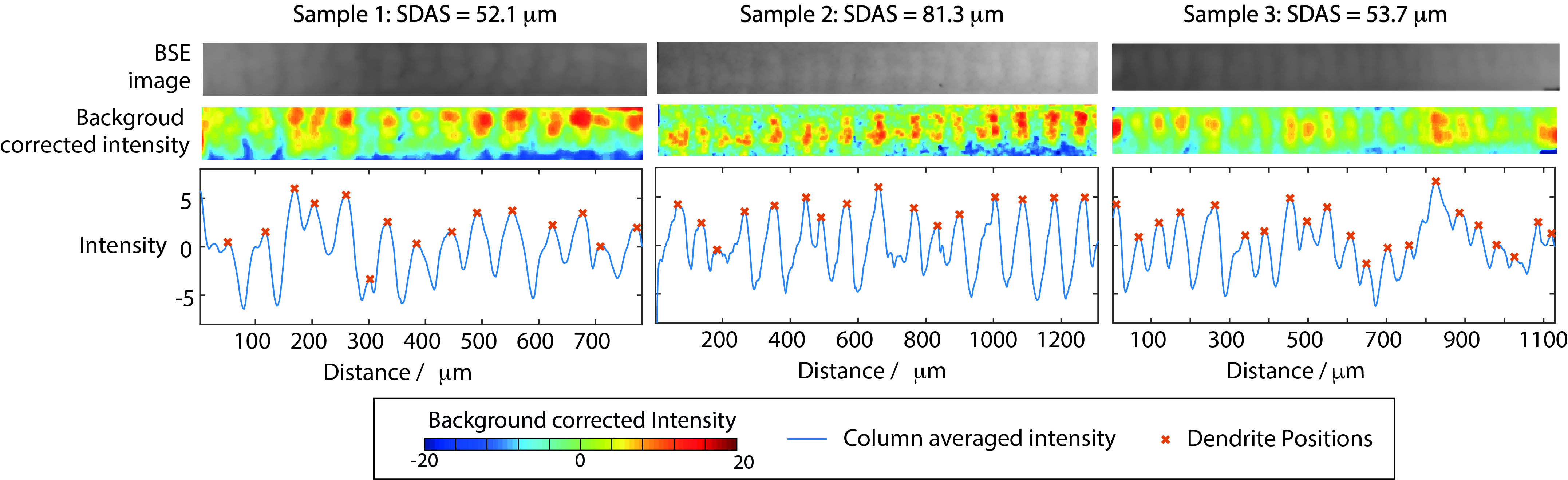}
		\caption{Measured secondary dendrite arm spacings calculated from BSE images of Samples 1, 2 \& 3.}
		\label{Figure_SDAS}
	\end{figure*}

Further differences between the microstructures of each sample is evident by measuring the secondary dendrite arm spacing (SDAS) observed in the gauge section. To measure the SDAS values in each sample, regions of the BSE images shown in Fig. \ref{Fig_microstructures} were cropped to show a single dendrite, as shown in the top row in Fig. \ref{Figure_SDAS}. To identify individual secondary dendrite arms, the BSE images were background corrected by subtracting a linear intensity across the image. From the orientation of the dendrites shown, the position of individual dendrites was obtained by plotting the column averaged intensity, as a function of distance, as shown in the bottom row. The local maxima in intensity correspond to the positions of individual dendrite arms, and were found using an automated peak searching function in MATLAB. The average SDAS for Samples 1-3 were measure as 52.1 $\pm$ 16.2\,$\micro$m, 81.3 $\pm$ 18.4\,$\micro$m $\&$ 53.7$\pm$ 12.7\,$\micro$m, receptively. The differences here must be composition influenced, as secondary dendrite arm geometries are determined by composition controlled coarsening.

\subsection{Quantitative analysis of geometrically necessary dislocation density \& residual elastic stresses}

\begin{figure*}
		\centering
		\includegraphics[width=\textwidth]{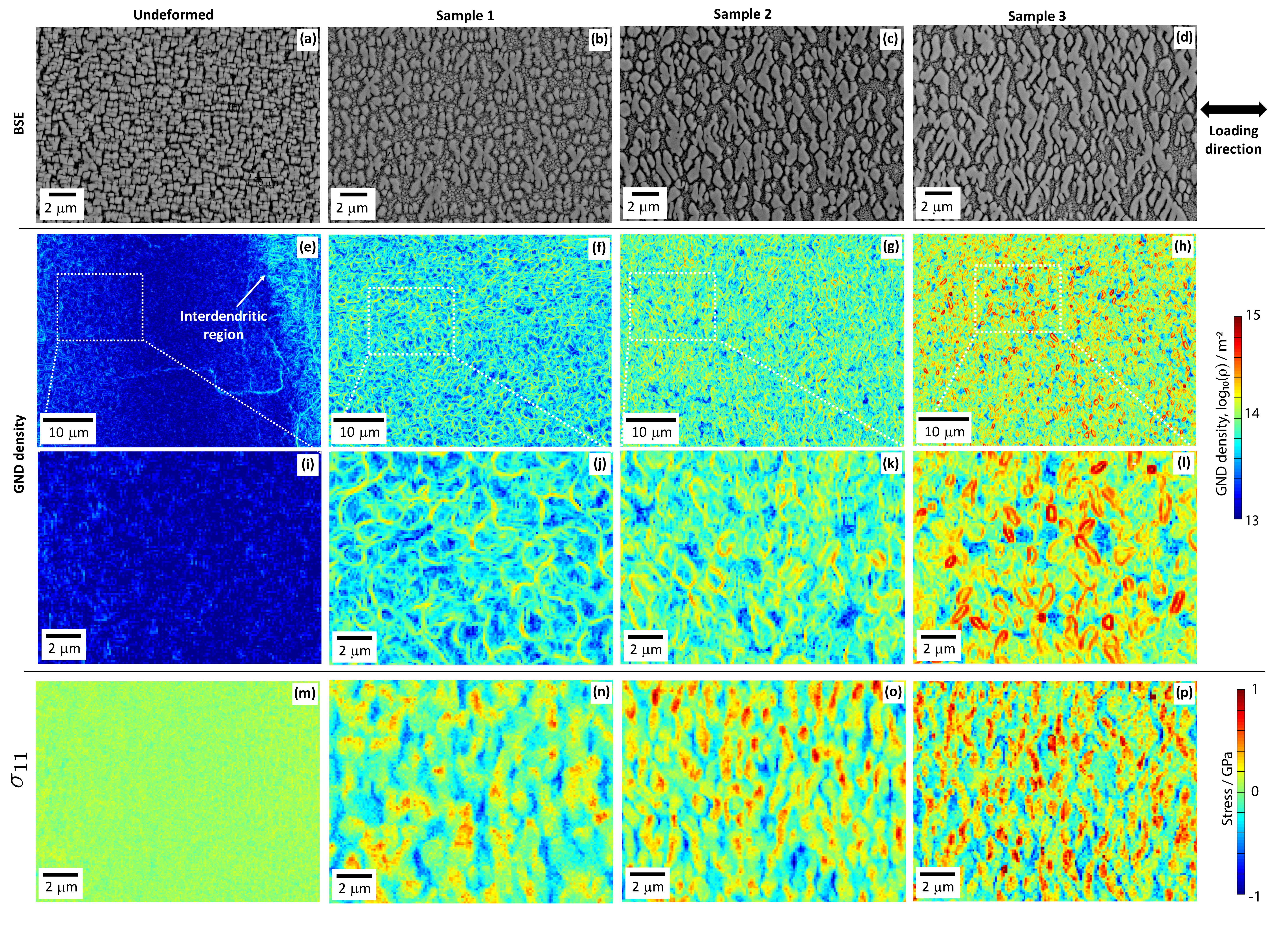}
		\caption{Overview of electron microscopy results obtained from each sample. SE micrographs of the microstructures are shown (a-d) with the $\gamma'$ precipitates within the darker contrast $\gamma$ matrix. A subset of the HR-EBSD maps are shown for each sample, including GND density distribution maps (e-h), and zoomed regions of each (i-l) to show dislocation structures. Finally, maps of one stress tensor component, $\sigma_{11}$ parallel to the loading direction, are shown.}
		\label{GND_summary}
	\end{figure*}

HR-EBSD was used to quantitatively determine the GND density and stress distributions following the ETMT process simulations, and benchmarked against a non-gauge section as a reference site. Micrographs of the different samples are shown in Fig. \ref{GND_summary} (a-d) where the $\gamma'$ morphologies are evident. The reference samples has a unimodal distribution of approximately cuboidal $\gamma'$, whereas samples following the process simulation have coarser primary $\gamma'$ and a fine dispersion of secondary $\gamma'$ between them. The primary $\gamma'$ have a characteristic `X' shape with highly directional protrusions, most obvious in Sample 1 Fig. \ref{GND_summary}, giving a topologically inverted $\gamma/\gamma'$ microstructure often observed following creep \cite{EPISHIN20014017}. Sample 3 has several $\gamma'$ precipitates that have coalesced in the direction transverse to the loading direction, characteristic of creep rafting e.g. \cite{MATAN19992031}.

Estimates of the GND densities of each sample are shown in Fig. \ref{GND_summary} (e-h). It is noted that a low GND density is evident in the dendrite core area in the non-gauge section whereas in the interdendritic region, arrowed in Fig. \ref{GND_summary}e for example, contained slightly coarser $\gamma'$ (see example in Fig. \ref{Fig_microstructures}g ) and a higher measured GND density. In contrast, the GND density in the gauge increased drastically for samples that underwent deformation, see Fig. \ref{GND_summary} (f-g). These region correspond to locations within the dendrite cores, chosen as they are known to be the sites for preferential recrystallisation \cite{Jo2003}. The magnified GND density maps in Fig. \ref{GND_summary} (i-l) show dislocation localisation at primary $\gamma/\gamma'$ interfaces. All samples have evidence of loop structures around individual primary $\gamma'$ precipitates, but is most clearly seen in Sample 3, which was subjected to the highest plastic strain. Orowan looping was evidently the mechanism by which plasticity was mediated.

Residual elastic stresses have been calculated from each sample, where the 3D stress tensor was obtained. As an example to illustrate the emergent patterns, maps of the $\sigma_{11}$ component (parallel to the [100] tensile direction) are shown in Fig. \ref{GND_summary} (m-p). The patterning is observed to follow the $\gamma/\gamma'$ microstructure, indicating significant stress partitioning of magnitudes reaching $\pm\sim400$\,MPa following the process simulation. 

\begin{figure}
		\centering
		\includegraphics[width=\columnwidth]{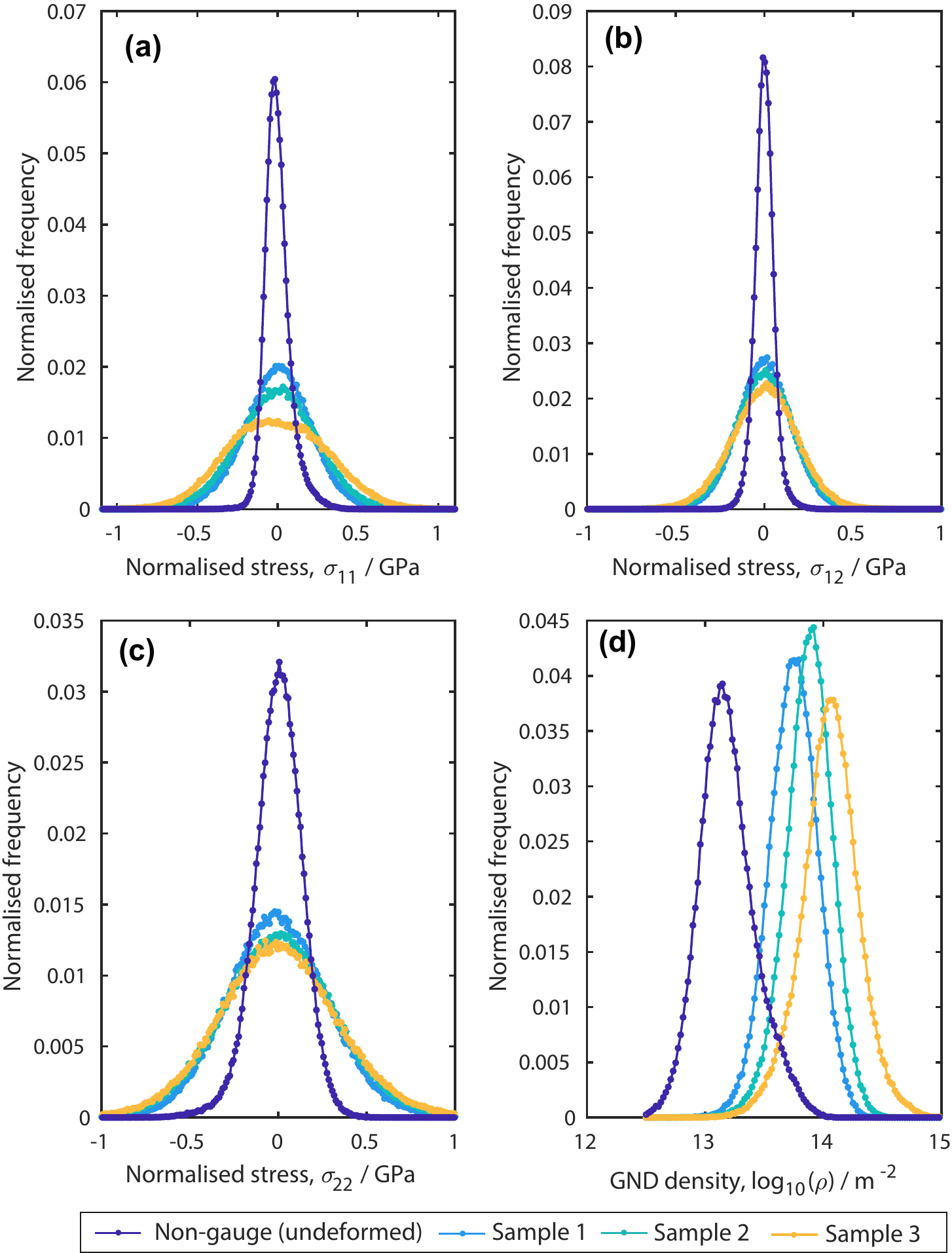}
		\caption{Histograms of HR-EBSD calculated stress tensor component distributions; (a) $\sigma_{11}$, (b) $\sigma_{12}$ \& (c) $\sigma_{22}$, and (d) GND density distributions. Each plot evaluates microstructures in the as-cast (Non-gauge, undeformed) state, and those that underwent process simulations (Samples 1-3).}
		\label{GND_histograms}
	\end{figure}

Quantitative comparisons of the in-plane stress tensor components and GND densities for each sample are evaluated as histograms in Fig. \ref{GND_histograms}. The distributions of the stress components $\sigma_{11}$, $\sigma_{12}$ \& $\sigma_{22}$, Fig. \ref{GND_histograms} (a-d), all have a Gaussian distribution and were subsequently fitted with this function; their widths are summarised in Table \ref{TableStresses}. The stress range of all tensor components increase relative to the undeformed condition following the process simulation. In summary, for all stress tensor components, $\sigma_{ij}$(Sample 1) $<$ $\sigma_{ij}$(Sample 2) $<$ $\sigma_{ij}$(Sample 3). These differences can be explained from their macroscopic behaviour; Samples 1 \& 2 reached the same plastic strain, but the latter reached a higher macroscopic tensile stress. Whilst Samples 2 \& 3 both reached $\sim100$\,MPa, the additional creep strain of Sample 3 resulted in increased residual elastic stresses. Notably for all samples, the transverse stresses $w_{\sigma_{22}}$, were higher than those in the tensile direction, $w_{\sigma_{11}}$, and significantly higher than the shear stresses, $w_{\sigma_{12}}$. The GND distributions, Fig. \ref{GND_histograms}d similarly had approximately Gaussian distributions with the exception of the undeformed sample which had a leading tail to skew the histogram, corresponding to the locally high GND density of the interdendritic region (Fig. \ref{GND_summary}e). The mean GND densities for each sample replicated the rank of residual stress magnitudes, $\rho$(Sample 1) $<$ $\rho$(Sample 2) $<$ $\rho$(Sample 3), as stated in Table \ref{TableStresses}.

\begin{table}[]
\centering
\caption{\label{TableStresses}For each sample, the stress tensor component distributions are evaluated, expressed here as the $w_{\sigma_{ij}}$, for the width of a fitted Gaussian function and its fitting error derived from the $\pm95\%$ confidence intervals. Similarly, each sample had its GND density distribution evaluated, expressed as the mean with the error representing the width of the fitted Gaussian function.}

\begin{tabular}{ccccc}
\hline \hline
                      & $w_{\sigma_{11}}$ & $w_{\sigma_{12}}$ & $w_{\sigma_{22}}$ & Mean log$_{10}(\rho)$ \\
                      \hline
                      & MPa                                  & MPa                                  & MPa                                  & m$^{-2}$       \\
                      \hline
Undeformed & $91\pm1$            & $68\pm1$            & $178\pm1 $          & $13.2\pm0.3 $    \\
Sample 1              & $285\pm1 $          & $212\pm1 $          &$ 399\pm2$           & $13.8\pm0.3  $   \\
Sample 2              & $339\pm1  $         & $230\pm1   $         &$ 443\pm2   $        & $13.9\pm0.3 $    \\
Sample 3              & $447\pm4  $         & $256\pm1   $         & $466\pm2  $        & $14.1\pm0.3 $ \\
\hline 
\end{tabular}
\end{table}

\subsection{Deformation patterning of individual precipitates}

\begin{figure*}
		\centering
		\includegraphics[width=110mm]{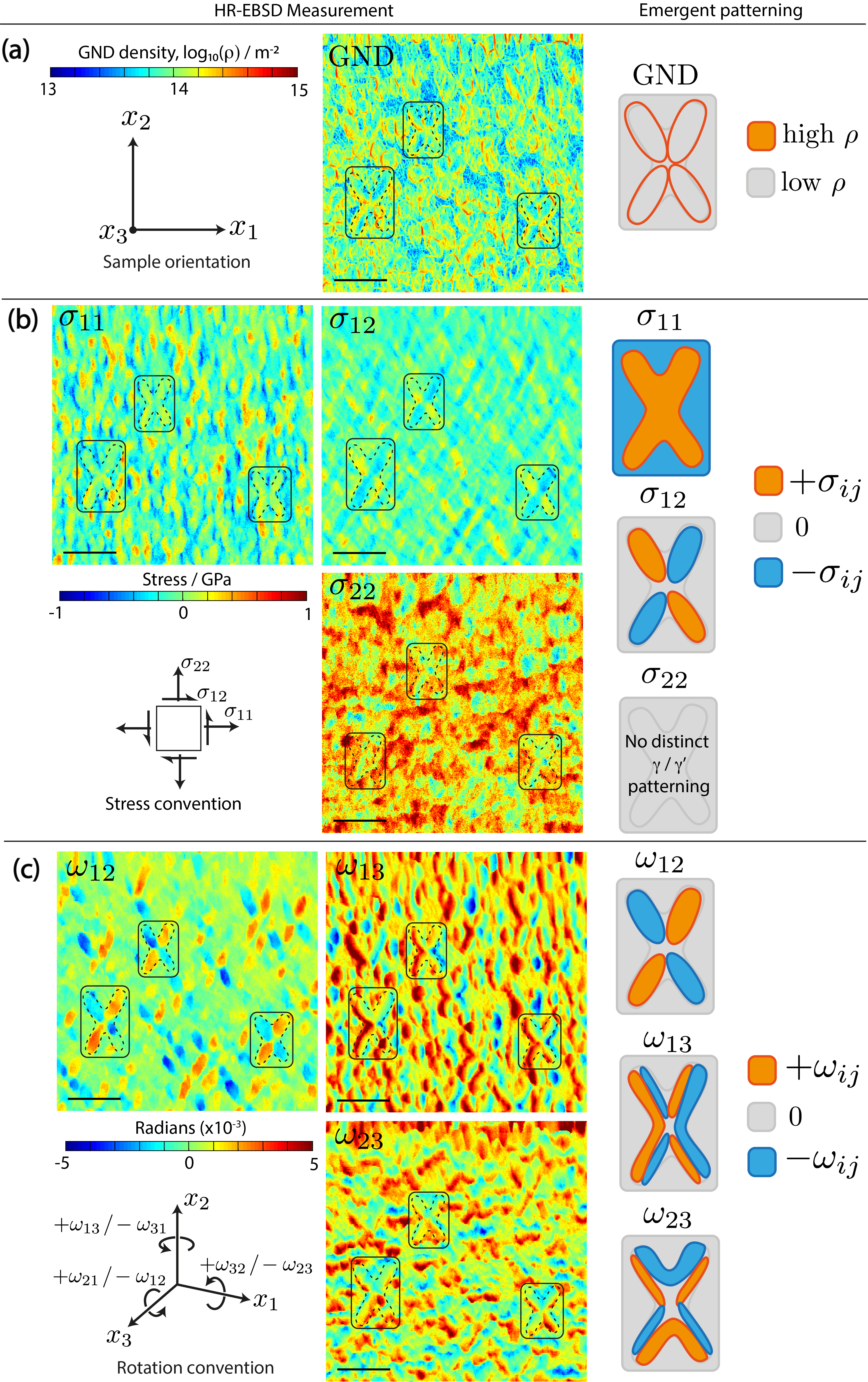}
		\caption{The deformation patterning from Sample 1 following its process simulation is assessed, considering (a) the GND density field, (b) the in-plane stress tensor components, and (c) the measured rotation tensor components. From the measured HR-EBSD datasets (left), approximations of the emergent patterns around individual $\gamma'$ precipitates are illustrated (right).}
		\label{Sample1_XEBSD}
	\end{figure*}

A subset of the HR-EBSD results have been studied in detail to identify the deformation patterning that emerges at the length scale of individual $\gamma'$ precipitates. A case study for this is shown in Fig. \ref{Sample1_XEBSD} showing measurements of (a) GND density, (b) in-plane stress tensor components, and (c) rotation tensor components. Illustrations that summarise the corresponding measurements are shown to the right. 

Each HR-EBSD map has an approximated outline of 3 $\gamma'$ precipitates, enabling comparison between each measured component. All maps including stress and rotation components are shown in the sample reference frame, denoted by the $x_1-x_2-x_3$ axes. In Fig. \ref{Sample1_XEBSD}a, it is evident that the $\gamma/\gamma'$ interfaces are the regions with the highest dislocation density; the presence of complete loops around individual $\gamma'$ arms is due to the 2D cross section though the morphologically complex $\gamma'$ precipitates. It is not believed that high dislocation densities exist within the individual $\gamma'$ precipitates. For the measured stress tensor components, Fig \ref{Sample1_XEBSD}a, $\sigma_{11}$ and $\sigma_{12}$ show characteristic patterning within the $\gamma'$ precipitates. The former is nominally tensile within the $\gamma'$ and compressive within the $\gamma$ matrix, whereas the shear stress, $\sigma_{12}$, is largely neutral within the $\gamma$ and alternating between +ve/-ve stresses in the $\gamma'$ protrusions. The $\sigma_{22}$ does not possess such clear stress partitioning, however, the stress sign is typically opposite to the $\sigma_{11}$; tensile in the $\gamma$ and compressive in the $\gamma'$ phase. 

Fig. \ref{Sample1_XEBSD}c shows rotation tensor components. Like the stress fields, the rotations including $\omega_{12}$, $\omega_{13}$ and $\omega_{23}$ possess distinctive patterns in the vicinity of $\gamma'$. The $\omega_{12}$ depicts +ve rotation in bottom-left and top-right $\gamma'$ protrusions and -ve for the other two. The $\omega_{13}$ and $\omega_{23}$ terms depict a +ve pattern on the left or bottom of the precipitates, whereas -ve pattern at the other half. 

The GND density, $\sigma_{ij}$ and $\omega_{ij}$ results obtained from this HR-EBSD analysis must develop from a number of factors including (1) the solute transport history that produce the microstructure, (2) the $\gamma'$ morphology shows the onset of topological inversion, (3) the anisotropic elastic properties of the $\gamma$ and $\gamma'$ phases,  and (4) their interaction with the imposed macroscopic deformation. These factors are considered in further detail in the following section.

\section{Discussion}

\subsection{Macroscopic Behaviour}

A variation in mechanical response including strain hardening rate and resultant macroscopic stress state, combined with a sharp strain hardening rate increase at the different $\gamma'$ formation temperatures indicate a significant influence of differing compositions between the samples. The origin of the dramatic sample-to-sample variation of such mm-scale specimens has been previously seen to result from significant micro-segregation in the cast bar from which these samples were machined \cite{DSOUZA2020138862}. This controls variation in local chemistry, influencing critical parameters such as $\gamma'$ solvus temperature, and consequently thermo-mechanical response. The DSC measurements independently verified this observation, indicating the transformation temperatures ($\gamma'$ solvus, eutectic and liquidus temperatures) can vary within a $\sim$10\,$^\circ$C range. 

\begin{figure}
		\centering
		\includegraphics[width=0.8\columnwidth]{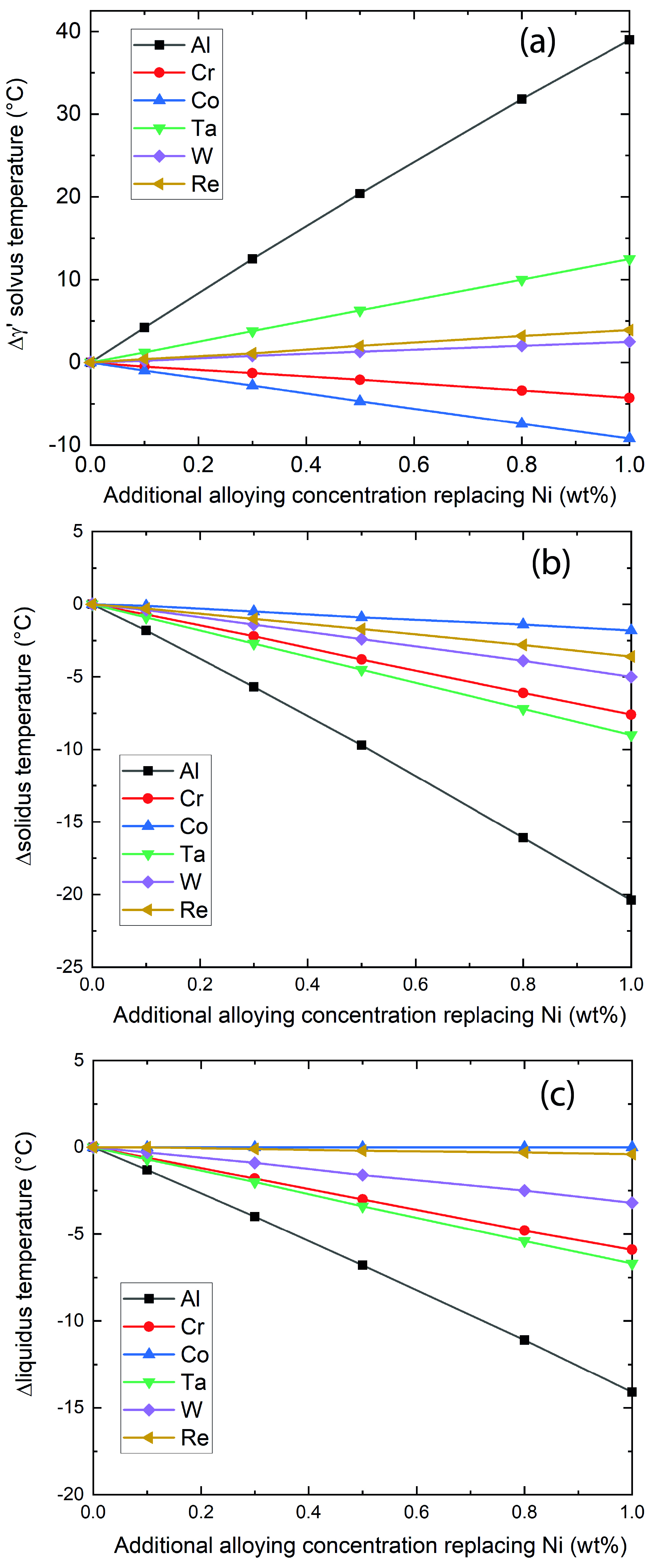}
		\caption{Simulated changes to phase transition temperatures by replacing alloy composition (up to 1\,wt.$\%$), approximating small concentration changes within the CMSX-4 alloy cast. }
		\label{Figure_ThermoCalc}
	\end{figure}

The {\it{Thermo-Calc}} software \cite{ANDERSSON2002273} using the {\it{TCNI8}} database \cite{TCNI8} was used to rationalise the sensitivity of changes in minor elements and their effect on the $\gamma'$ solvus, solidus and liquidus temperatures. Principal alloying elements were altered by replacing Ni, up to 1\,wt$\%$, with different elements. By making these small changes, the transformation temperatures including the $\gamma'$ solvus, solidus and liquid temperatures are predicted to change by several degrees Celsius, as shown in Fig. \ref{Figure_ThermoCalc}. Clearly, each element has the potential to alter these transformation temperatures to different extents, but are most influenced by the $\gamma'$ forming elements. Of the three DSC measurements made (Fig. \ref{Figure_DSC}), the sample with the highest $\gamma'$ solvus had lowest liquidus temperature; this sample behaviour is consistent with the {\it{Thermo-Calc}} measurements where a higher content of $\gamma'$ forming elements (i.e. Al and Ta) must be present. From the mechanical data obtained from samples in this study, Sample 2, was deemed to have the highest $\gamma'$ solvus temperature, indicating an enrichment in these elements. The earlier formation of strengthening $\gamma'$ precipitates during cooling gives rise to a much higher resultant macroscopic stress, as the material work hardens over a greater time/strain, over Sample 1, which had a lower $\gamma'$ solvus temperature. This is consistent with the greater resulting residual elastic stresses and GND density.

In addition to composition differences, other features from the sample will influence the macroscopic response. This includes the presence of casting pores, which are expected to be created via the coalescence of cast-in vacancies plus those created by the Kirkendall vacancy generation mechanism, which form due to the solidification induced composition gradients \cite{ANTON1985173}.  These appeared elongated following deformation, where their presence and distribution could effect the macroscopic behaviour. Furthermore, differences between the SDAS was observed between samples. Whilst there is a known correlation between the superalloy cooling rate and the SDAS \cite{Milenkovic2012216,NAWROCKI2020153,ZHANG2010889}, this is unlikely to be the reason for the differences observed. Again, the effect of local composition must also be the controlling factor; elements such as Cr, Co, Re and Re are known to segregate to dendrite core regions, whilst the $\gamma'$ forming elements have greater affinity to interdendritic regions \cite{Karunaratne2000,Ganesan2005} and the magnitude of the partitioning coefficients of these elements in particular have a strong correlation to the SDAS \cite{Matache2016}. Notably, as the SDAS increases, the partitioning coefficient of W and Re increases, whilst they lowers for the $\gamma'$ forming elements. Considering significant micro-segregation in the cast bars of CMSX-4 has been previously measured to vary over length scale of hundreds of microns \cite{DSOUZA2020138862}, of similar dimensions to the gauge lengths of the samples tested here, the SDAS spacing and consequently the macroscopic behaviour differences are inevitable.

The effect of casting pores, SDAS between samples, as well as the presence of additional features of the microstructure including eutectic phases and TCP phases could each influence the macroscopic behaviour of the alloy during the process simulation. However, the strong correlation between the stress-strain behaviour and the resultant residual elastic stresses and GND density (recall Fig. \ref{GND_histograms}) indicate the behaviour is largely dominated by the deformation interaction of the $\gamma/\gamma'$ microstructure within the dendrite cores. Upon cooling, Sample 1 and Sample 2 had different $\gamma'$ solvus temperatures, indicating their dissimilar compositions, as was evidence from the onset of increased work hardening in each sample as the precipitates formed. The precipitates in Sample 2 formed first, resulting in a longer plastic strain range where the material work hardening, resulting in a higher residual stress and GND density.

\subsection{Evolution of $\gamma'$ morphology}

 	\begin{figure*}
		\centering
		\includegraphics[width=0.9\textwidth]{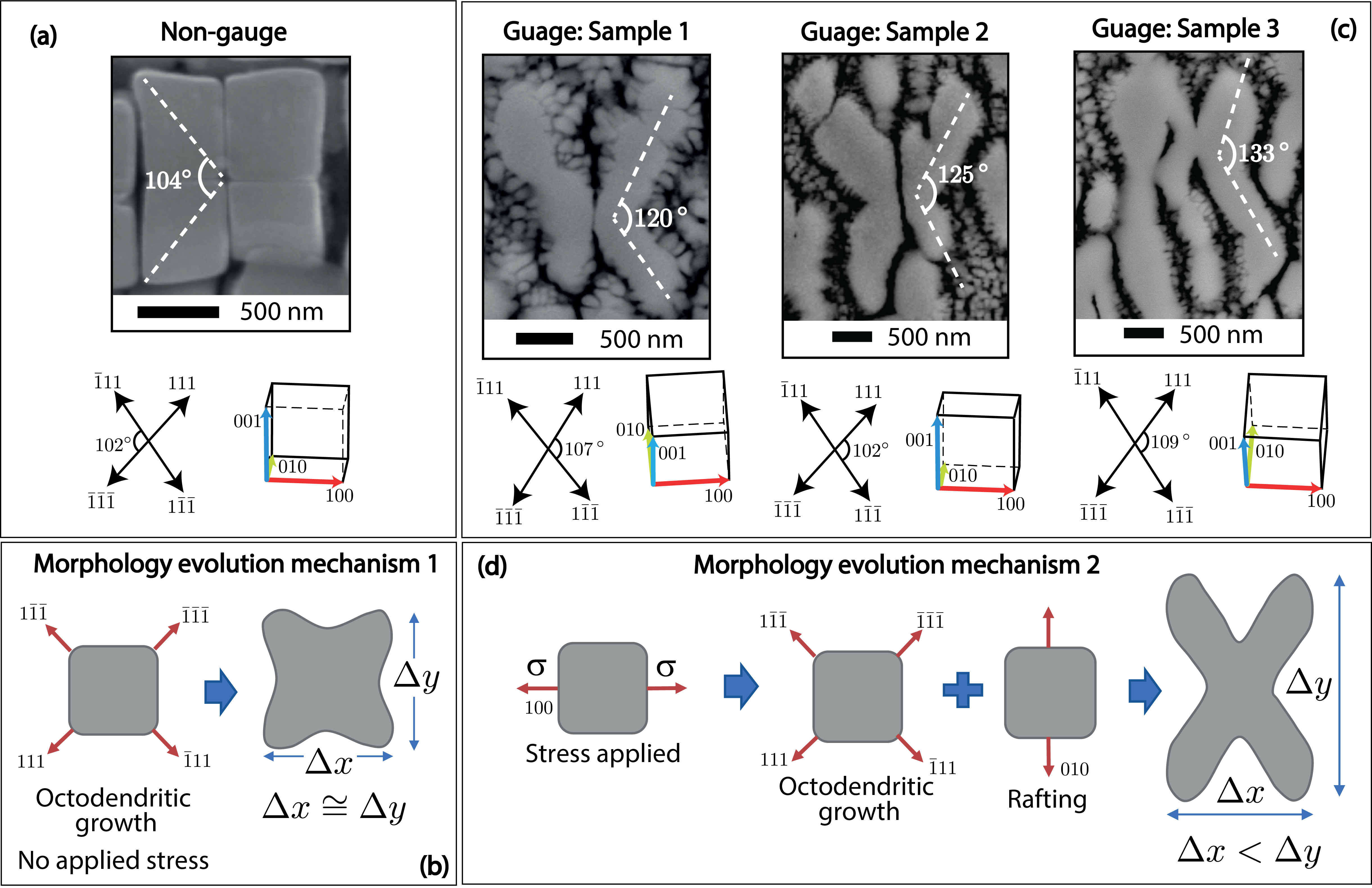}
		\caption{Example $\gamma'$ precipitate morphologies observed in (a) the non-gauge, reference material, and (b) in the gauge regions of Samples 1, 2 \& 3 after the ETMT process simulation. Illustrations of the corresponding growth mechanisms are shown in (c) \& (d), respectively.}
		\label{Figure_GP_shape}
	\end{figure*}
	
The morphology of the $\gamma'$ precipitates following the process simulation are rationalised here.  The general features of the $\gamma'$ morphologies in the deformed and undeformed states are summarised in Fig. \ref{Figure_GP_shape}. The $\gamma'$ in the as-cast state has an approximately cuboidal structure with small protrusions at the corners, Fig. \ref{Figure_GP_shape}a, indicating growth along the body diagonals. This observation is commonly termed octodendritic growth \cite{Ricks1983}, where the $\gamma'$ precipitates possess a preferential $\langle$111$\rangle$  growth direction, as illustrated by Fig. \ref{Figure_GP_shape}b. This growth mechanism was confirmed by measuring a near identical angle between the precipitate diagonals and the angle between $\langle111\rangle$ directions, traced onto the plane of the sample surface. This same method was applied to the micrographs of the deformed specimens (Fig. \ref{Figure_GP_shape}c), however, the angle between the apparent growth directions (parallel to the length of the $\gamma'$ protrusions) was greater than the angle between the trace of the $\langle111\rangle$ directions, indicating an additional mechanism was operative. 

For single crystal nickel-base superalloys studied after elevated temperature deformation, often in a creep regime, the accumulated plasticity at elevated temperatures results in a highly rafted $\gamma/\gamma'$ microstructure \cite{MATAN19992031} and dislocations that are confined to the $\gamma$ channels. Rafting, in particular for negatively misfitting alloys, such as for CMSX-4 studied here, rafting is associated with the elongation of precipitates in the direction transverse to loading (N-type) \cite{POLLOCK19941859, Tien1972}. The `X' shape $\gamma'$ morphology that has preferential elongation in the direction transverse to loading indicates a joint mechanism is operative; this includes the simultaneous action of octodendritic growth and N-type rafting. Sample 1 and Sample 2, which reached the same level of strain have a similar $\gamma'$ morphology, whereas Sample 3 which was an additionally crept has a less regular $\gamma'$ morphology (recall Fig. \ref{GND_summary}), with several precipitates coalescing along the direction transverse to loading; rafting has evidently dominated the final microstructure.
	
The `X' morphology $\gamma'$ precipitates additionally have a fine dispersion of $\gamma'$ precipitates at the interface,  formed preferentially in a region of high dislocation density. As these $\gamma'$ are spherical and small, these are likely to have formed during the final rapid cooling (at 5\,$\circ$C/sec), when no macroscopic load was applied. These interfacial locations would be favourable for heterogeneous nucleation, plus any solutes segregation to the dislocations themselves would promote precipitate growth. Likewise, the `X' morphology $\gamma'$ precipitates required significant solute transport to become highly inverted from its initially cuboidal form. A recently studied system with an inverted $\gamma/\gamma'$ microstructure after creep found that the microstructure was assisted by the transport of solute via plasticity assisted redistribution, which includes $\gamma'$ shearing dislocations to assist the transport of solutes to and from a $\gamma/\gamma'$ interface \cite{ANTONOV2020287}. No $\gamma'$ shearing was evident in this case, however, significant plasticity along the length of protrusions, at the $\gamma/\gamma'$ interfaces, would provide an obvious mechanism to assist further growth along the precipitate diagonals.

		 \begin{figure}
		\centering
		\includegraphics[width=0.8\columnwidth]{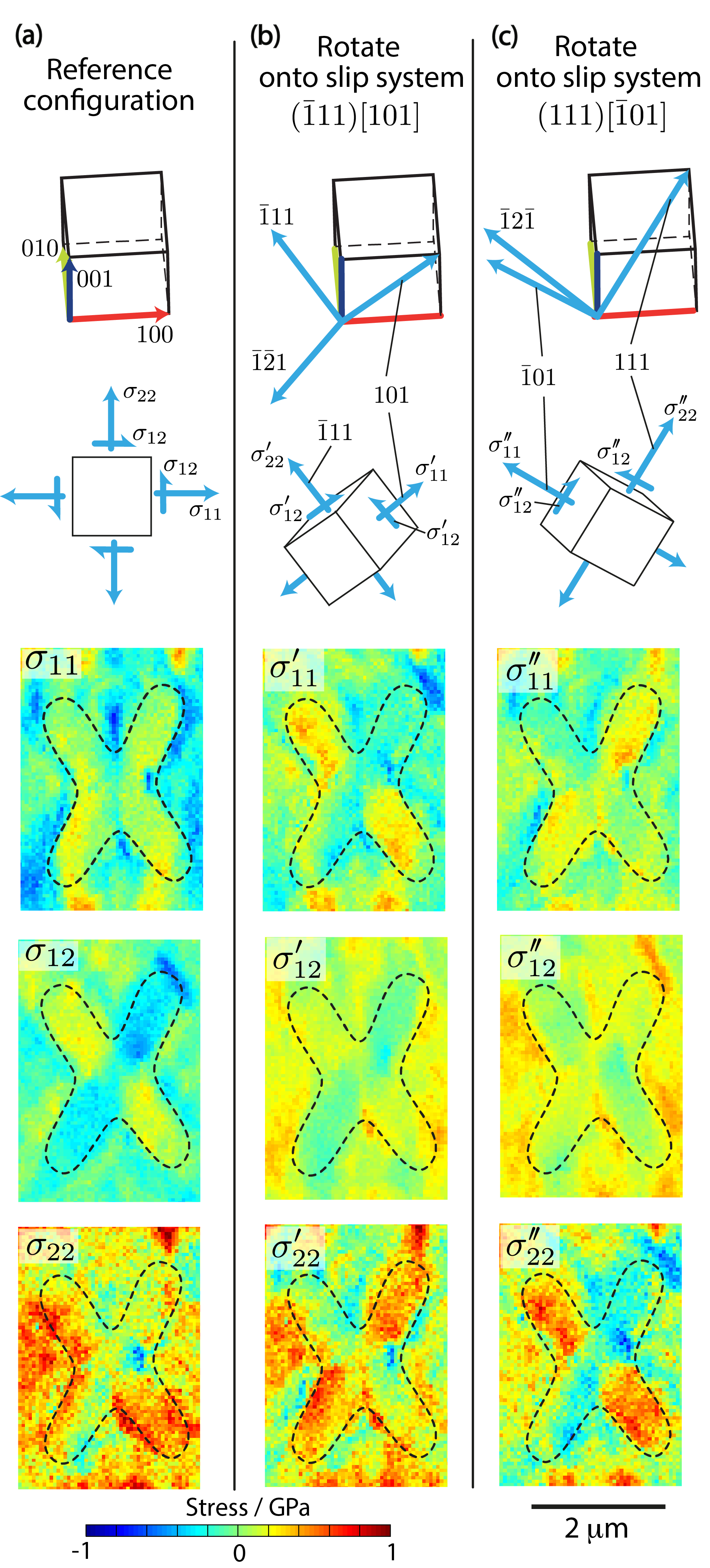}
		\caption{(a) Selected in-plane tensor components of the residual elastic stress fields in the sample references frame, then rotated into (b) the $(\bar{1}11)[101]$ slip system, and (c) the $(111)[\bar{1}01]$ slip system. Note, the $\sigma_{11}'$ \& $\sigma_{11}''$ terms are parallel to the slip directions, and $\sigma_{22}'$ \& $\sigma_{22}''$ are parallel to the slip plane normals, in (b) and (c) respectively.}
		\label{Figure_RotatedStresses}
	\end{figure}

\subsection{Shape of stress fields}

The HR-EBSD measured stress patterns are complex, where it is difficult to separate the mechanical and thermodynamics contributions to the resultant precipitate shape. Instead of considering their patterning in the sample reference frame, they have been rotated into orientations that would most favoured slip. From the [100] loading direction (which is also parallel to the [100]) crystal direction, two example slip systems, assuming slip was of type $\{111\}\langle\bar{1}10\rangle$ (e.g. REF 123), with the highest Schmid factor were considered further. The stress tensor components were rotated such that $\sigma_{11}$ was set parallel to the slip direction and $\sigma_{22}$ was parallel to the slip plane normal. The stress transformation procedure adopted here is summarised the Appendix. For a general description of such crystallographic transformations, the reader is referred elsewhere \cite{BRITTON2016113}. 

The rotated stress tensor HR-EBSD maps for a selected region from Sample 1 containing a single $\gamma'$ precipitate are shown in Fig. \ref{Figure_RotatedStresses}. The in-plane stress tensor components are shown with respect to the reference configuration (stresses in the sample reference frame, equivalent to Fig. \ref{GND_summary} and Fig. \ref{Sample1_XEBSD}) is shown in Fig. \ref{Figure_RotatedStresses}a, into an orientation to summarises the $(\bar{1}11)[101]$ slip system, Fig. \ref{Figure_RotatedStresses}b, and the $(111)[\bar{1}01]$ slip systems, Fig. \ref{Figure_RotatedStresses}c. The axes of the evaluated orientations are shown in the top row, and the corresponding stress components shown beneath. 

As expected, the emergent patterns in each $\sigma_{ij}$ component in the reference configuration are shown to be dissimilar to rotated stresses. The notable result is that $\gamma'$ protrusions are slightly compressive along the body diagonals (close the the $\langle111\rangle$ directions), and tensile in the direction transverse to the $\gamma'$ protrusion (close the the $\langle\bar{1}10\rangle$ directions). This implies that as the $\gamma'$ protrusions grow outwards, the elastic strain along body diagonals should reduce, making this growth pattern favourable. The sign of these normal stresses also implies that along the length of the protrusions, the lattice misfit is negative, but must flip sign at the $\gamma'$ protrusion tips, making the lattice misfit locally positive. For this $\gamma'$ with an elaboated morphology in conjunction with a high elastic strain gradient can only be accommodated by interfacial dislocations, observed as regions of high density GNDs (recall Fig. \ref{Sample1_XEBSD}a). Such equilibrium interfacial dislocation networks are often observed in crept samples \cite{Gabb1989} at early stages of deformation, though is often not considered to possess Orowan looping \cite{Field1992} as is evident here.
	
The $\sigma_{12}$ shear stress map possesses a pattern of $\pm$200\,MPa within each $\gamma'$ protrusion, as shown in Fig. \ref{Figure_RotatedStresses}a, however, in the rotated orientations, the $\sigma_{12}$ stress has a magnitude close to zero within the $\gamma'$ precipitates, and larger stresses within the $\gamma$ channels, close to the $\gamma/\gamma'$ interfaces. With limited stress acting on the slip plane within the $\gamma'$ precipitates, it is clear why a low GND density was observed within them. If this sample were to be loaded, one would expected the $\gamma$ matrix, having high magnitude shear stresses resolved onto the slip planes to yield first. This would prevail at elevated temperature if the lattice misfit strains remain present. Whilst elevated temperature deformation studies have often directly observed dislocation shearing of $\gamma'$ precipitates via dislocation ribbons \cite{Kear1973, LEONCAZARES202047} or indirectly from stress relaxation measurements, \cite{COLLINS2017103}, the resultant stress state does not favour this. It is proposed that this unique stress state is a consequence of the precipitate morphology, where the rafting influenced $\gamma'$ growth is balanced by retaining coherency and thus accumulates significant normal stresses along the length and transverse to the protrusions. The combination of normal stresses imposed on these protrusions give a maximum shear stress within the $\gamma'$ precipitates in a direction close to the reference axes, $\langle100\rangle$, in an orientation where slip is not permitted. Conversely, a minimal shear stress acting on a slip plane, $\langle111\rangle$, where plasticity could prevail, results in no plasticity with the $\gamma'$ precipitates. This is a significant factor is constraining plasticity to the $\gamma$ and $\gamma/\gamma'$ interfaces only.

\subsection{Implication to Recrystallisation}

A combination of factors that will undoubtedly contribute to the susceptibility of a single crystal nickel-base superalloy to recrystallisation have been revealed. If one applied the observations from macro-specimens to a real turbine blade cast, one can speculate the following; (1) location specific macroscopic deformation is caused by micro-segregation, which give rise to (2) differing hardening rates and resultant stress levels reached within the casting, where stress levels are higher for regions with a greater concentration of $\gamma'$ forming elements. Furthermore, (3) the applied stress combined with the cooling from ultra high temperatures will control the morphology of the inverted $\gamma/\gamma'$ microstructure, where (4) the morphology of the $\gamma'$ precipitates themselves result in a stress state that prohibits any $\gamma'$ shearing, and promotes dislocation formation in the $\gamma$ channels and at the $\gamma/\gamma'$ interfaces. Greater accumulation of these dislocation structures will be expected close to geometric features where stress concentrations are created. This will be enhanced by the thermal mismatch between the metal and ceramic core plus the presence of high magnitude residual elastic stresses at the microstructure length scale. As a result, the recrystallisation propensity during post processing homogenisation heat treatments will be enhanced.

\section{Summary and Conclusions}

A meso-scale thermal-mechanical process simulation on the single crystal nickel-base superalloy CMSX-4 was conducted to understand the microstructure and deformation patterning that develops during an industrial casting process. As-cast material was cooled from a near solidus temperature whilst subjecting samples to controlled tensile loads, and understood using a combination of macroscopic measurements and mortem electron microscopy including HR-EBSD characterisation. The following specific conclusions can be drawn from this work:

\begin{enumerate}

\item{During sample cooling, strain hardening rates varied significantly between samples; prevalent in both the high temperature $\gamma$ phase field and at lower temperatures once $\gamma'$ precipitates had formed. Evidence of different $\gamma'$ solvus temperatures and secondary dendrite arm spacings between samples indicates significant composition segregation and microstructural heterogeneity that must also be present in turbine blade castings.}

\item{Resultant microstructures following the process simulation comprise partially transformed eutectic with preferential $\langle 111 \rangle$ directionality, casting pores elongated in the tensile direction, and $\gamma'$ precipitates with a characteristic `X' morphology. This morphology is rationalised by considering two simultaneously active mechanisms; solute transport via N-type rafting and lattice misfit -- elastic anisotropic interactions that favour octodendritic growth.}

\item{HR-EBSD measurements within dendrite cores reveal significant deformation patterning that follow the $\gamma/\gamma'$ microstructure. GND density is increased with increased plastic strains, with dislocations localised to $\gamma/\gamma'$ interfaces, with fewer present within $\gamma'$ precipitates or $\gamma$ channels. The pinned structures evidently reduce the $\gamma/\gamma'$ load partitioning as residual elastic stresses significantly increase.}

\item{Inspection of individual $\gamma'$ precipitates revealed no precipitate cutting, instead plasticity is mediated via Orowan looping. The stress tensor rotated into an orientation where slip is likely to be favoured indicates the `X' $\gamma'$ precipitate morphology heavily exacerbates the stress heterogeneity; a significant build up of normal stresses and minimal shear stress results, explaining why dislocations are pinned to $\gamma/\gamma'$interfaces.}

\item{Samples that developed the most significant residual stresses, found in microstructure regions that overlap and neighbour fields of  high localised dislocation density are considered to be representative of locations within a turbine blade cast that would be most susceptible to the recrystallisation defect in post casting homogenisation heat treatments.}

\end{enumerate}

\section*{Acknowledgments}

The authors would like to acknowledge the provision of material from Rolls-Royce. D.C. acknowledges financial support from his Birmingham Fellowship and C.P. \& Y.T. would like to acknowledge the funding from the Engineering and Physical Science Research Council (EPSRC) under UKRI Innovation Fellowship EP/S000828/1.

\section*{References}


\begin{thebibliography}{10}
\expandafter\ifx\csname url\endcsname\relax
  \def\url#1{\texttt{#1}}\fi
\expandafter\ifx\csname urlprefix\endcsname\relax\def\urlprefix{URL }\fi
\expandafter\ifx\csname href\endcsname\relax
  \def\href#1#2{#2} \def\path#1{#1}\fi

\bibitem{ReedSuperalloys}
{R.C. Reed}, The Superalloys: Fundamentals and Applications, Cambridge
  University Press, 2006.

\bibitem{Dobbs1996}
J.~Dobbs, J.~Graves, S.~Meshkov, Advanced airfoil fabrication, in:
  R.~Kissinger, D.~Deye, D.~Anton, A.~Cetel, M.~N. abd T.M.~Pollock,
  D.~Woodford (Eds.), Superalloys 1996, The Minerals, Metals and Materials
  Society, Warrendale, PA, 1996, p. 263?272.

\bibitem{LI20151}
{Z. Li and J. Xiong and Q. Xu and J. Li and B. Liu}, Deformation and
  recrystallization of single crystal nickel-based superalloys during
  investment casting, J. Mater. Process Tech. 217 (2015) 1 -- 12.

\bibitem{Burgel2000}
{J. Meng and T. Jin and X.F. Sun and Z.Q. Hu}, Recrystallization in single
  crystals of nickel base superalloys, in: T.~Pollock, R.~Kissinger, R.~Bowman,
  K.~Green, M.~McLean, S.Olson, J.~Schirra (Eds.), Superalloys 2000, The
  Minerals, Metals and Materials Society, Warrendale, PA, 2000, p. 229?238.

\bibitem{Meng2010}
{J. Meng and T. Jin and X.F. Sun and Z.Q. Hu}, Effect of surface
  recrystallization on the creep rupture properties of a nickel-base single
  crystal superalloy, Mater. Sci. Eng. A 527 (2010) 6119--6122.

\bibitem{Moverare2009}
{J.J. Moverare and S. Johansson and R.C. Reed}, Deformation and damage
  mechanisms during thermal-mechanical fatigue of a single-crystal superalloy,
  Acta Mater. 57 (2009) 2266--2276.

\bibitem{Cox2003}
{D.C. Cox and B. Roebuck and C.M.F. Rae and R.C. Reed}, {Recrystallisation of
  single crystal superalloy CMSX-4}, Mater. Sci. Tech. 19 (2003) 440--446.

\bibitem{Panwisawas2013}
{C. Panwisawas and H. Mathur and J.-C. Gebelin and D. Putman and C.M.F. Rae and
  R.C. Reed}, {Prediction of recrystallization in investment cast
  single-crystal superalloys}, Acta Mater. 61 (2013) 51--66.

\bibitem{Porter1981}
{A. Porter and B. Ralph}, {The recrystallization of nickel-base superalloys},
  J. Mater. Sci. 16 (1981) 707--713.

\bibitem{Dahlen1980}
{M. Dahl\'en and L. Einberg}, {The influence of $\gamma'$-precipitation on the
  recrystallization of a nickel base superalloy}, Acta Metall. 28 (1980)
  41--50.

\bibitem{Wang2012}
{L. Wang and F. Pyczak and J. Zhang and L.H. Lou, R.F. Singer}, {Effect of
  eutectics on plastic strain deformation and subsequent recrystallization in
  the single crystal nickel base superalloy CMSX-4}, Mater. Sci. Eng. A 532
  (2012) 487--492.

\bibitem{Mathur2014}
{H.N. Mathur and C.N. Jones and C.M.F. Rae}, {A study on the effect of
  composition, and the mechanisms of recrystallization in single crystal
  Ni-based superalloys}, MATEC Web Conf. 14 (2017) 7003.

\bibitem{Jo2003}
{C.-Y. Jo and H.-Y. Cho and H.-M. Kim}, {Effect of recrystallisation on
  microstructural evolution and mechanical properties of single crystal nickel
  base superalloy CMSX-2 Part 1 - Microstructural evolution during
  recrystallisation of single crystal}, Mater. Sci. Tech 19 (2003) 1665--1670.

\bibitem{DUPEUX19872203}
{M. Dupeux and J. Henriet and M. Ignat}, {Tensile stress relaxation behaviour
  of Ni-based superalloy single crystals between 973 and 1273 K}, Acta Metall.
  35 (1987) 2203 -- 2212.

\bibitem{WANG20131179}
{H.Wang and B.Clausen and C.N. Tom\'{e} and P.D. Wu}, Studying the effect of
  stress relaxation and creep on lattice strain evolution of stainless steel
  under tension, Acta Mater. 61 (2013) 1179 -- 1188.

\bibitem{SVOBODA1997125}
{J. Svoboda and P. Luk\'{a}\v{s}}, Modelling of recovery controlled creep in
  nickel-base superalloy single crystals, Acta Mater. 45 (1997) 125 -- 135.

\bibitem{PREUNER2009973}
{J. Preu{\ss}ner and Y. Rudnik and H. Brehm and R. V\"olkl and U. Glatzel}, A
  dislocation density based material model to simulate the anisotropic creep
  behavior of single-phase and two-phase single crystals, Int. J. Plast. 25
  (2009) 973 -- 994.

\bibitem{Giraud2013}
{R. Giraud and Z. Hervier and J. Cormier and G. Saint-Martin and F. Hamon and
  X. Milhet and J. Mendez}, Strain effect on the $\gamma'$ dissolution at high
  temperatures of a nickel-based single crystal superalloy, Metall. Mater.
  Trans. A 44 (2013) 131 -- 146.

\bibitem{Schwalbe2018}
{C. Schwalbe and J. Cormier and C.N. Jones, E. Galindo-Nava and C.M.F. Rae},
  Investigating the dislocation-driven micro-mechanical response under
  non-isothermal creep conditions in single-crystal superalloys, Metall. Mater.
  Trans. A 49 (2018) 3988 -- 4002.

\bibitem{CORMIER20076250}
{J. Cormier and X. Milhet and J. Mendez}, {Non-isothermal creep at very high
  temperature of the nickel-based single crystal superalloy MC2}, Acta Mater.
  55 (2007) 6250 -- 6259.

\bibitem{CORMIER20106300}
{J. Cormier and G. Cailletaud}, Constitutive modeling of the creep behavior of
  single crystal superalloys under non-isothermal conditions inducing phase
  transformations, Mater. Sci. Eng. A 527 (2010) 6300 -- 6312.

\bibitem{Cormier2010}
{J. Cormier and M. Jouiad and F. Hamon and P. Villechaise and X. Milhet}, Very
  high temperature creep behavior of a single crystal ni-based superalloy under
  complex thermal cycling conditions, Phil. Mag. Lett. 90 (2010) 611--620.

\bibitem{LEGRAVEREND201455}
{J.-B. {le Graverend} and J. Cormier and F. Gallerneau and P. Villechaise and
  S. Kruch and J. Mendez}, A microstructure-sensitive constitutive modeling of
  the inelastic behavior of single crystal nickel-based superalloys at very
  high temperature, Int. J. Plast. 59 (2014) 55 -- 83.

\bibitem{LEGRAVEREND2014990}
{J.-B. {le Graverend} and J. Cormier and F. Gallerneau and S. Kruch and J.
  Mendez}, Highly non-linear creep life induced by a short close
  $\gamma'$-solvus overheating and a prior microstructure degradation on a
  nickel-based single crystal superalloy, Mater. Des. 56 (2014) 990 -- 997.

\bibitem{MA20081657}
{A. Ma and D. Dye and R.C. Reed}, {A model for the creep deformation behaviour
  of single-crystal superalloy CMSX-4}, Acta Mater. 56 (2008) 1657 -- 1670.

\bibitem{ZHU20124888}
{Z. Zhu and H. Basoalto and N. Warnken and R.C. Reed}, A model for the creep
  deformation behaviour of nickel-based single crystal superalloys, Acta Mater.
  60 (2012) 4888 -- 4900.

\bibitem{Panwisawas2017}
{C. Panwisawas and N. D'Souza and D.M. Collins and A. Bhowmik}, The contrasting
  roles of creep and stress relaxation in the time-dependent deformation during
  in-situ cooling of a nickel-base single crystal superalloy, Sci. Rep. 7
  (2017) 11145.

\bibitem{Panwisawas2018}
{C. Panwisawas and N. D'Souza and D.M. Collins and A. Bhowmik and B. Roebuck},
  History dependence of the microstructure on time-dependent deformation during
  in-situ cooling of a nickel-based single-crystal superalloy, Metall. Mater.
  Trans. A 49 (2018) 3963--3972.

\bibitem{DSOUZA2020138862}
{N. D'Souza and B. Roebuck and D.M. Collins and G.D. West and C. Panwisawas},
  Relating micro-segregation to site specific high temperature deformation in
  single crystal nickel-base superalloy castings, Mater. Sci. Eng. A 773 (2020)
  138862.

\bibitem{Roebuck2001}
{B. Roebuck and D.C. Cox and R.C. Reed}, {The temperature dependence of gamma
  prime volume fraction in a Ni-base single crystal superalloy from resistivity
  measurements}, Scripta Mater. 44 (2001) 917 -- 921.

\bibitem{Roebuck2004}
{B. Roebuck and D.C. Cox and R.C. Reed}, An innovative device for the
  mechanical testing of miniature specimens of superalloys, in: {K.A. Green and
  T.M. Pollock and H. Harada and T.E. Howson and R.C. Reed and J.J. Schirra,
  and S. Walston} (Ed.), Superalloys 2004, The Minerals, Metals and Materials
  Society, Warrendale, PA, 2004, pp. 523--528.

\bibitem{Sulzer2004}
{S. Sulzer and E. Alabort and A. N\'emeth and B. Roebuck and R. Reed}, {On the
  rapid assessment of mechanical behaviour of a prototype Ni-based superalloy
  using small-scale testing}, Metall. Mater. Trans. A 49 (2018) 4214 -- 4235.

\bibitem{WILKINSON2006307}
{A.J. Wilkinson and G. Meaden and D.J. Dingley}, {High-resolution elastic
  strain measurement from electron backscatter diffraction patterns: New levels
  of sensitivity}, Ultramicroscopy 106 (2006) 307 -- 313.

\bibitem{BRITTON20111395}
{T.B. Britton and A.J. Wilkinson}, Measurement of residual elastic strain and
  lattice rotations with high resolution electron backscatter diffraction,
  Ultramicroscopy 111 (2011) 1395 -- 1404.

\bibitem{Villert2009}
{S. Villert and C. Maurice and C. Wyon and R. Fortunier}, {Accuracy assessment
  of elastic strain measurement by EBSD}, J. Microsc. 233 (2009) 290--301.

\bibitem{Britton_2018}
{T.B. Britton and J.L.R. Hickey}, Understanding deformation with high angular
  resolution electron backscatter diffraction ({HR}-{EBSD}), {IOP} Conference
  Series: Materials Science and Engineering 304 (2018) 012003.

\bibitem{WILKINSON2012366}
{A.J. Wilkinson and T.B. Britton}, {Strains, planes, and EBSD in materials
  science}, Mater. Today 15 (2012) 366 -- 376.

\bibitem{Britton2013}
{B. Britton, I. Holton, G. Meaden and D. Dingley}, High angular resolution
  electron backscatter diffraction: measurement of strain in functional and
  structural materials, Microscopy and Analysis 27 (2013) 8 -- 13.

\bibitem{WMG2006_1}
{A.J. Wilkinson and G. Meaden and D.J. Dingley}, High resolution mapping of
  strains and rotations using electron backscatter diffraction, Mater. Sci
  Tech. 22 (2006) 1271--1278.

\bibitem{ARSENLIS19991597}
{A. Arsenlis and D.M Parks}, Crystallographic aspects of
  geometrically-necessary and statistically-stored dislocation density, Acta
  Mater. 47 (1999) 1597 -- 1611.

\bibitem{Nye}
{J. Nye}, Some geometrical relations in dislocated crystals, Acta Metall. 1
  (1953) 153 -- 162.

\bibitem{PANTLEON2008994}
{W. Pantleon}, Resolving the geometrically necessary dislocation content by
  conventional electron backscattering diffraction, Scripta Mater. 58 (2008)
  994 -- 997.

\bibitem{Wilkinson2010}
{A.J. Wilkinson and D. Randman}, Determination of elastic strain fields and
  geometrically necessary dislocation distributions near nanoindents using
  electron back scatter diffraction, Phil. Mag. 90 (2010) 1159--1177.

\bibitem{Fuchs2001}
{G.E. Fuchs}, {Solution Heat Treatment Response of a Third Generation Single
  Crystal Ni-base Superalloy}, Mater. Sci. Eng. A 300 (2001) 52--60.

\bibitem{Wilson2003}
{B.C. Wilson and J.A. Hickman and G.E. Fuchs}, {The effect of solution heat
  treatment on a single-crystal Ni-based superalloy}, J. Mater. 55 (2003)
  35--40.

\bibitem{EPISHIN20014017}
{A. Epishin and T. Link and U. Br\"uckner and P.D. Portella}, Kinetics of the
  topological inversion of the $\gamma/\gamma'$ -microstructure during creep of
  a nickel-based superalloy, Acta Mater. 49 (2001) 4017 -- 4023.

\bibitem{MATAN19992031}
{N. Matan and D.C. Cox and C.M.F. Rae and R.C. Reed}, {On the kinetics of
  rafting in CMSX-4 superalloy single crystals}, Acta Mater. 47 (1999) 2031 --
  2045.

\bibitem{ANDERSSON2002273}
{J-O Andersson and T. Helander and L. H\"oglund and P. Shi and B. Sundman},
  {Thermo-Calc \& DICTRA, computational tools for materials science}, Calphad
  26 (2002) 273 -- 312.

\bibitem{TCNI8}
{TCNI8 TCS Ni-based Superalloys Database, v8.1} ({Accessed April 2020}).

\bibitem{ANTON1985173}
{D.L. Anton and A.F. Giamei}, Porosity distribution and growth during
  homogenization in single crystals of a nickel-base superalloy, Mater. Sci.
  Eng. 76 (1985) 173 -- 180.

\bibitem{Milenkovic2012216}
{Milenkovic, S. and Sabirov, I. and Llorca, J.}, {Effect of the cooling rate on
  microstructure and hardness of MAR-M247 Ni-based superalloy}, Mater. Lett. 73
  (2012) 216--219.

\bibitem{NAWROCKI2020153}
{J. Nawrocki and M. Motyka and D. Szeliga and W. Ziaja and R. Cygan and J.
  Sieniawski}, Effect of cooling rate on macro- and microstructure of
  thin-walled nickel superalloy precision castings, J. Manuf. Processes 49
  (2020) 153 -- 161.

\bibitem{ZHANG2010889}
{J. Zhang and J. Li and T. Jin and X. Sun and Z. Hu}, {Effect of solidification
  parameters on the microstructure and creep property of a single crystal
  Ni-base superalloy}, J. Mater. Sci. Tech. 26~(10) (2010) 889 -- 894.

\bibitem{Karunaratne2000}
{D.C. Cox and P. Carter and R.C. Reed}, {Modelling of the microsegregation in
  CMSX-4 Superalloy and its homogenization during Heat Treatment}, in: {T.M.
  Pollock and R.D. Kissinger and R.R. Bowman and K.A. Green and M. McLean and
  S.Olson and J.J. Schirra} (Ed.), Superalloys 2000, The Minerals, Metals and
  Materials Society, Warrendale, PA, 2000, p. 263?272.

\bibitem{Ganesan2005}
{M. Ganesan and D. Dye and P.D. Lee}, A technique for characterizing
  microsegregation in multicomponent alloys and its application to
  single-crystal superalloy castings, Metall. Mat. Trans. A 36 (2005)
  2191--2204.

\bibitem{Matache2016}
{G. Matache and D. M. Stefanescu and C. Puscasu and E. Alexandrescu},
  {Dendritic segregation and arm spacing in directionally solidified CMSX-4
  superalloy}, Int. J. Cast Metals Research 29 (2016) 303--316.

\bibitem{Ricks1983}
{Ricks, R. and Porter, A.J. and Ecob, R.C.}, {The Growth of $\gamma'$
  Precipitates in Nickel-Base Superalloy}, Acta Metall. 31 (1983) 43--53.

\bibitem{POLLOCK19941859}
{T.M. Pollock and A.S. Argon}, Directional coarsening in nickel-base single
  crystals with high volume fractions of coherent precipitates, Acta Metall.
  Mater. 42 (1994) 1859 -- 1874.

\bibitem{Tien1972}
{J.K. Tien and R.P. Gamble}, Effects of stress coarsening on coherent particle,
  Metall. Trans. 3 (1972) 2157 -- 2162.

\bibitem{ANTONOV2020287}
{S. Antonov and Y. Zheng and J.M. Sosa and H.L. Fraser and J. Cormier and P.
  Kontis and B. Gault}, Plasticity assisted redistribution of solutes leading
  to topological inversion during creep of superalloys, Scripta Mater. 186
  (2020) 287 -- 292.

\bibitem{BRITTON2016113}
{T.B. Britton and J. Jiang and Y. Guo and A. Vilalta-Clemente and D. Wallis and
  L.N. Hansen and A. Winkelmann and A.J. Wilkinson}, Tutorial: Crystal
  orientations and ebsd - or which way is up?, Materials Characterization 117
  (2016) 113 -- 126.

\bibitem{Gabb1989}
{T.P. Gabb and D.L. Draper and D.R. Hull and R.A. MacKay and M.V. Nathal}, The
  role of interfacial dislocation networks in high temperature creep of
  superalloys, Mater. Sci. Eng. A 118 (1989) 59 -- 69.

\bibitem{Field1992}
{R.D. Field and T.M. Pollock and W.H. Murphy}, {The development of
  $\gamma/\gamma'$ interfacial dislocation networks during creep in Ni-base
  superalloys}, in: {S.D. Antolovich and R.W. Stusrud and R.A. MacKay and D.L.
  Anton and T. Khan and R.D. Kissinger and D.L. Klarstrom} (Ed.), Superalloys
  1992, The Minerals, Metals and Materials Society, Warrendale, PA, 1992, pp.
  557--566.

\bibitem{Kear1973}
{G.R. Leverant and B.H. Kear}, The mechanism of creep in gamma prime
  precipitation hardened nickel-base alloys at intermediate temperatures,
  Metall. Trans. 1 (1973) 491 -- 498.

\bibitem{LEONCAZARES202047}
{F.D. Le\'{o}n-C\'{a}zares and R. Schl\"{u}tter and T. Jackson and E.I.
  Galindo-Nava and C.M.F. Rae}, A multiscale study on the morphology and
  evolution of slip bands in a nickel-based superalloy during low cycle
  fatigue, Acta Mater. 182 (2020) 47 -- 59.

\bibitem{COLLINS2017103}
{D.M. Collins and N. D'Souza and C. Panwisawas}, In-situ neutron diffraction
  during stress relaxation of a single crystal nickel-base superalloy, Scripta
  Mater. 131 (2017) 103 -- 107.

\end{thebibliography}

\section{Appendix}

The crystal orientation is defined by a rotation of the Cartesian basis vectors $X$, $Y$ and $Z$, which is described by a series of rotation matrix operations with rotations in each defined by Euler angles. Following the `ZXZ' Bunge convention, the operation is 

\begin{equation}
    {\bf{R}} = {\bf{R}}_{Z(\phi_2)}{\bf{R}}_{X(\Phi)}{\bf{R}}_{Z(\phi_1)}
\end{equation}

\noindent where each rotation matrix represents a mathematically positive rotation about a reference direction: 

\begin{equation}
   {\bf{R}}_Z =  \begin{pmatrix}
\cos \theta & -\sin\theta & 0\\
\sin\theta & \cos\theta & 0\\
0 & 0 & 1
\end{pmatrix}
\end{equation}

\noindent  and

\begin{equation}
   {\bf{R}}_X = \begin{pmatrix}
1 & 0 & 0 \\
0 & \cos \theta & -\sin\theta\\
0 & \sin\theta & \cos\theta

\end{pmatrix}
\end{equation}

\noindent where the angle $\theta$ may be replaced by the Euler angles, $\phi_1$, $\Phi$ or $\phi_2$. In this successive transformation operation, $\phi_1$ is first rotated about $Z$, then $\Phi$ about the transformed X axis, then, $\phi_2$ about the new  Z axis. The rotated basis vectors from the crystal frame ${\bf{u}}_{\rm c}$ can be related to the sample frame, ${\bf{u}}_{\rm s}$, via

\begin{equation}
    {\bf{u}}_{\rm s} = {\bf{u}}_{\rm c}{\bf{R}}
\end{equation}

\noindent The rotated configuration for the desired slip system requires vectors in the sample frame that are congruent to the slip direction and slip plane normal. Both the slip plane normal, ${\bf{n}}_{\rm c}$, and slip direction, ${\bf{s}}_{\rm c}$, in the crystal frame comprise crystallographic indices, $uvw$. In the sample reference frame:

\begin{equation}
    {\bf{n}}_{\rm s} = {\bf{n}}_{\rm c}{\bf{R}}
\end{equation}

\noindent and

\begin{equation}
    {\bf{s}}_{\rm s} = {\bf{s}}_{\rm c}{\bf{R}}
\end{equation}

\noindent A right handed basis set is completed via:

\begin{equation}
    \bf{p}_{\rm s} = \bf{n}_{\rm s} \times \bf{s}_{\rm s}
\end{equation}

\noindent where $\bf{p}_{\rm s}$ is a direction that will lie in the slip plane. The vectors ${\bf{n}}_{\rm s}$, ${\bf{s}}_{\rm s}$ and ${\bf{p}}_{\rm s}$ are next used to create a rotation matrix, ${\bf{R}}_{\rm ss}$, that represents the orientation of the slip system:

\begin{equation}
    {\bf{R}}_{\rm ss} = \hat{\bf{s}}^{\rm T}_{\rm s}[1\, 0\, 0] + \hat{\bf{n}}^{\rm T}_{\rm s}[0\, 1\, 0] + \hat{\bf{p}}^{\rm T}_{\rm s}[0\, 0\, 1]
\end{equation}

\noindent where the superscript $\rm T$ is the transpose. This rotation matrix is finally used to rotate the stress tensor,

\begin{equation}
    \boldsymbol{\sigma}' = {\bf{R}}_{\rm ss} \boldsymbol{\sigma} {\bf{R}}_{\rm ss}^{\rm{T}}
\end{equation}

\noindent where $\sigma_{11}'$, $\sigma_{22}'$, $\sigma_{33}'$ each correspond to the slip direction $(\hat{\bf{s}})$, the slip plane normal $\hat{\bf{n}}$, and their orthogonal vector lying in the slip plane, $\hat{\bf{p}}$, respectively. When applied to an EBSD map, the above operation is performed independently for every discrete map data point, which have corresponding Euler angles and stress tensor, $\boldsymbol{\sigma}$.



\end{document}